\documentclass[letterpaper]{article} 
\usepackage{aaai24}  
\usepackage{times}  
\usepackage{helvet}  
\usepackage{courier}  
\usepackage[hyphens]{url}  
\usepackage{graphicx} 
\usepackage{natbib}  
\usepackage{caption} 
\usepackage{algorithmic}
\usepackage[linesnumbered,ruled,vlined]{algorithm2e}
\title{QCS-SGM+: Improved Quantized Compressed Sensing with Score-Based Generative Models}
\author {
    Xiangming Meng\textsuperscript{\rm 1}\thanks{Corresponding author.},
    Yoshiyuki Kabashima\textsuperscript{\rm 2}
}
\affiliations {
    \textsuperscript{\rm 1}The Zhejiang University-University of Illinois Urbana-Champaign Institute, Zhejiang University, China\\
    \textsuperscript{\rm 2}Institute for Physics of Intelligence \& Department of Physics, The University of Tokyo, Japan\\
    xiangmingmeng@intl.zju.edu.cn, kaba@phys.s.u-tokyo.ac.jp
}

\usepackage{bibentry}

\usepackage{amsfonts}
\usepackage{verbatim}
\usepackage{times}
\usepackage{amsmath}
\usepackage{multirow}
\usepackage{multicol}
\usepackage{kotex}
\usepackage{amssymb}
\usepackage{pifont}
\usepackage{subfigure}
\usepackage{subfig}
\usepackage{epstopdf}
\usepackage{array}
\usepackage{booktabs}

\SetKwInput{KwInput}{Input}                
\SetKwInput{KwInitialize}{Initialization}                
\SetKwInput{KwOutput}{Output}              
\usepackage{subfigure}
\usepackage{graphicx}
\usepackage{amsmath}

\begin{document}

\maketitle

\begin{abstract}

In practical compressed sensing (CS), the obtained measurements typically necessitate quantization to a limited number of bits prior to transmission or storage. This nonlinear quantization process poses significant recovery challenges, particularly with extreme coarse quantization such as 1-bit. Recently, an efficient algorithm called QCS-SGM was proposed for quantized CS (QCS) which utilizes score-based generative models (SGM) as an implicit prior. Due to the adeptness of SGM in capturing the intricate structures of natural signals, QCS-SGM substantially outperforms previous QCS methods. However, QCS-SGM is constrained to (approximately) row-orthogonal sensing matrices as the computation of the likelihood score becomes intractable otherwise. To address this limitation, we introduce an advanced variant of QCS-SGM, termed QCS-SGM+, capable of handling general matrices effectively. The key idea is a Bayesian inference perspective on the likelihood score computation, wherein expectation propagation is employed for its approximate computation.   Extensive experiments are conducted, demonstrating the substantial superiority of QCS-SGM+ over QCS-SGM for general sensing matrices beyond mere row-orthogonality.

\end{abstract}

\section{Introduction}
Compressed sensing (CS) has emerged as a ubiquitous paradigm in signal processing and machine learning, aiming to accurately reconstruct high-dimensional signals from a limited number of measurements \cite{donoho2006compressed,candes2008introduction}. The success of CS hinges on the assumption that, while the target signal may be high-dimensional, it possesses an inherent low-dimensional representation such as sparsity or low-rankness. Conventional CS models typically assume direct access to analog (continuous) measurements with infinite precision. In practice, however, analog measurements must be quantized into a finite number of digital bits using an analog-to-digital converter (ADC or quantizer) before further transmission, storage, or processing can occur \citep{boufounos2008one,zymnis2009compressed, dai2011information}. In the extreme case, measurements are quantized into a single bit, retaining only the sign information while disregarding the magnitude \cite{boufounos2008one}. Interestingly, 1-bit CS have gained attention due to their simplicity in hardware implementation and robustness against multiplicative errors \cite{boufounos2008one,zymnis2009compressed}. Nevertheless, such nonlinear quantization operation invariably leads to information loss, thereby undermining the effectiveness of standard CS algorithms designed for analog measurements. To address this issue, a plethora of  quantized CS (QCS) algorithms have been  developed by explicitly considering the quantization effect \citep{boufounos2008one,zymnis2009compressed, dai2011information, plan2012robust, plan2013one, jacques2013robust, Xu_2013, Xu_2014, awasthi2016learning,meng2018unified, jung2021quantized,liu2020sample,liu2022non, zhu2022unitary, meng2022quantized}. 
Among these, the recently proposed  algorithm QCS-SGM \cite{meng2022quantized} demonstrates exceptional state-of-the-art (SOTA) reconstruction performance under low-precision quantization levels. The key idea of QCS-SGM  lies in utilizing the powerful score-based generative models (SGM, also known as diffusion models) \citep{song2019generative, song2020improved, sohl2015deep, ho2020denoising,nichol2021improved} as an implicit prior for the target signal. Intuitively, from a Bayesian perspective, the more accurate the prior obtained, the fewer observations required. Owing to SGM's ability to capture the  intricate structure of the target signal $\bf{x}$ beyond simple sparsity, QCS-SGM can accurately reconstruct $\bf{x}$ even from a small number of severely quantized noisy measurements. 

While QCS-SGM exhibits remarkable performance, it has one fundamental limitation: it is derived under the assumption that the sensing matrix $\bf{A}$ is (approximately) row-orthogonal. For general matrices, QCS-SGM's performance will deteriorate since its computation of the \textit{likelihood score} (defined  in (\ref{bayes-rule-score})) becomes less accurate \citep{meng2022quantized}, as illustrated in Section \ref{sec:relation}. Although row-orthogonal matrices are
prevalent in conventional CS, in practical applications one might encounter other types of matrices due to non-ideal physical constraints or design choices.  In fact, the investigation of general sensing matrices  has long been an active and important topic, such as the popular ill-conditioned matrices and correlated matrices, among others \citep{manoel2015swept, schniter2016vector, tanaka2018performance, ihara2018typical, rangan2019vector,venkataramanan2022estimation, zhu2022unitary, fan2022approximate}. Despite these advances, much of existing work concentrates on standard CS or QCS with handcrafted sparsity. The study of QCS using SGM (diffusion models) for general sensing matrices, however, still remains a largely untouched research area. This paper try to address this problem and the main contributions are summarized as follows:
\begin{itemize}
\item We propose an advanced variant of QCS-SGM, designated as QCS-SGM+, which addresses the inherent limitation of the original QCS-SGM. Specifically, by treating the likelihood score computation as a Bayesian inference problem, QCS-SGM+ utilizes the well-established expectation propagation (EP) algorithm \citep{minka2001expectation} to yield a more refined approximation of the otherwise intractable likelihood score for general matrices. 

\item We validate the effectiveness of the proposed QCS-SGM+  in various experimental settings, encompassing diverse real-world datasets, distinct general matrices, and different noise levels. In each scenario, QCS-SGM+ consistently demonstrates remarkably superior performance over existing methods. 
\end{itemize}

\section{Related Work}
Quantized compressed sensing (QCS) was first introduced in \citet{boufounos2008one,zymnis2009compressed}, after which it has become an important research topic and various QCS algorithms have been proposed, including some theoretical analysis \citep{dai2011information, plan2012robust, plan2013one, jacques2013robust, Xu_2013, Xu_2014, awasthi2016learning,meng2018unified, jung2021quantized,liu2020sample,liu2022non, zhu2022unitary}. With the recent advent of deep generative models \citep{goodfellow2014generative,kingma2013auto, rezende2015variational, song2019generative, song2020improved, sohl2015deep, ho2020denoising,nichol2021improved}, there has been a rising interest in CS methods with data-driven priors \citep{bora2017compressed, hand2019global, asim2020invertible, pan2021exploiting, meng2022diffusion}. Specifically, following the pioneering CSGM framework \citep{bora2017compressed}, the  prior $p(\bf{x})$ of $\bf{x}$ is learned through a generative model, such as VAE  \citep{kingma2013auto},  GAN \citep{goodfellow2014generative}, and score-based generative models (SGM) or diffusion models (DM) \citep{song2019generative, song2020improved, sohl2015deep, ho2020denoising,nichol2021improved}. In the case of QCS, \citet{liu2020sample, liu2022non} extended the CSGM framework to non-linear observations such as 1-bit CS using VAE and GAN (in particular DCGAN \citep{radford2015unsupervised}).  Surprisingly, SGM or DM \citep{song2019generative, song2020improved, ho2020denoising,nichol2021improved}  
have demonstrated superior effectiveness, even surpassing state-of-the-art GAN \citep{goodfellow2014generative} and VAE \citep{kingma2013auto} in generating diverse natural sources. In line with this, \citet{meng2022quantized} recently proposed a novel algorithm, QCS-SGM, employing SGM as an implicit prior, achieving state-of-the-art reconstruction performances for QCS. However, its application remains confined to (approximate) row-orthogonal sensing matrices. 

\section{Preliminary}
\subsection{System Model}
The problem of quantized CS (QCS) can be mathematically formulated as follows \citep{boufounos2008one,zymnis2009compressed}
\begin{align}
    \bf{y} = \mathsf{Q}(\bf{Ax+n}), \label{quantized model}
\end{align}
where the goal is to recover  an  unknown high-dimensional signal ${\bf{x}}\in \mathbb{R} ^{N\times 1}$ from a set of quantized measurements ${\bf{y}}\in \mathbb{R} ^{M\times 1}$, where ${\bf{A}} \in \mathbb{R} ^{M \times N}$ is a known linear sensing matrix,  ${\bf{n}}\sim \mathcal{N}({\bf{n}};0,\sigma^2 {\bf{I}})$  is an i.i.d. additive Gaussian noise, and $\mathsf{Q}(\cdot): \mathbb{R}^{M \times 1} \to \mathcal{Q}^{M \times 1} $ is an \textit{element-wise} quantizer function which maps each element into a finite (or countable) set of codewords $\mathcal{Q}$, i.e., $ y_m= \mathsf{Q}{(z_m+n_m)}\in \mathcal{Q}$, or equivalently $(z_m+n_m) \in \mathsf{Q}^{-1}{(y_m)},m=1,2,...,M$, where $z_m$ is the $m$-th element of  $\bf{z}={\bf{Ax}}$.  Same as \citet{meng2022quantized}, we consider the uniform quantizer with $Q$ quantization bits (resolution). The  quantization codewords $ \mathcal{Q}=\{q_r\}_{r=1}^{2^{Q}}$ consist of $2^{Q}$ elements, each with a quantization interval $\mathsf{Q}^{-1}{(q_r)}=[l_{q_r},u_{q_r})$, where $l_{q_r}$ and $u_{q_r}$are the lower and upper quantization threshold associated with  the codeword $q_r$. 
In the extreme 1-bit case, i.e., $Q = 1$,  only the signs are observed so that (\ref{quantized model}) reduces to
\begin{align}
    \bf{y} = \rm{sign}(\bf{Ax+n}), \label{1-bit model}
\end{align}
which corresponds to the well-known 1-bit CS and the quantization codewords are $\mathcal{Q}=\{-1,+1\}$. 
\subsection{QCS-SGM: Quantized CS with SGM}
Compared to standard CS without quantization, QCS is particularly more challenging due to two key factors: (1) quantization leads to information loss, especially with low quantization resolution; (2) the nonlinearity of quantization operations can cause standard CS algorithms to deteriorate  when applied directly. Recently, inspired by the prowess of SCM \citep{song2019generative,song2020score} in density estimation, \citet{meng2022quantized} proposed an efficient method called QCS-SGM for QCS which can accurately reconstruct the target signal from a small number of severely quantized noisy measurements. The basic idea of QCS-SGM is to perform posterior sampling from $p({\bf{x}}\mid {\bf{y}})$ by using a learned SGM  as an implicit prior \citep{meng2022quantized}. Specifically, by utilizing the annealed Langevin dynamics (ALD) \citep{song2020score}, the posterior samples can be iteratively obtained as follows 
\begin{align}
    {\bf{x}}_{t} = {\bf{x}}_{t-1} + \alpha_t \nabla_{{\bf{x}}_{t-1}} \log{p({\bf{x}}_{t-1}\mid {\bf{y}})} +\sqrt{2\alpha_t} {\bf{z}}_t,\; 1\leq t\leq T \label{Langevin-dynamics-posterior},
\end{align}
where the conditional (\textit{posterior}) score $\nabla_{{\bf{x}}_{t}} \log{p({\bf{x}}_{t}\mid {\bf{y}})}$ is required. Using the Bayesian rule, the $\nabla_{{\bf{x}}_{t}} \log{p({\bf{x}}_{t}\mid {\bf{y}})}$ is decomposed into two terms
\begin{align}
   \underbrace{\nabla_{{\bf{x}}_{t}} \log{p({\bf{x}}_{t}\mid {\bf{y}})}}_{\textit{posterior score}}   =  \underbrace{\nabla_{{\bf{x}}_t} \log{p({\bf{x}}_t)}}_{\textit{prior score}}  +  \underbrace{\nabla_{{\bf{x}}_t} \log{p({\bf{y}}\mid {\bf{x}}_t})}_{\textit{likelihood score}},\label{bayes-rule-score}
\end{align}
including  the unconditional score $ \nabla_{{\bf{x}}_t} \log{p({\bf{x}}_t)}$ (called \textit{ prior score} in \citet{meng2022quantized}), and the conditional score $\nabla_{{\bf{x}}_t} \log{p({\bf{y}}\mid {\bf{x}}_t})$ (called  \textit{likelihood score} in \citet{meng2022quantized}). While the prior score $\nabla_{{\bf{x}}_t} \log{p({\bf{x}}_t)}$  can be readily computed   
using a pre-trained score network, the likelihood score $\nabla_{{\bf{x}}_t} \log{p({\bf{y}}\mid {\bf{x}}_t})$ is generally intractable. To circumvent this difficulty, \citet{meng2022quantized} proposed a simple yet effective approximation of $\nabla_{{\bf{x}}_t} \log{p({\bf{y}}\mid {\bf{x}}_t})$ under an uninformative prior assumption, whereby ${\bf{x}}_t$ is approximated as ${\bf{x}}_t = {\bf{x}} + \beta_t \tilde{\bold{n}}$, where $\tilde{\bold{n}}\sim \mathcal{N}({\bf{0}}, \bold{I})$. As a result, substituting it into (\ref{quantized model}), we obtain an equivalent representation as  
\begin{align}
    {\bf{y}} = \mathsf{Q}\left({\bf{A}} {\bf{x}}_t + \tilde{\bf{n}}_t  \right), \label{quantized-model-x-tilde}
\end{align}
where  $\tilde{{\bf{n}}}_t\sim \mathcal{N}({\bf{0}},\sigma^2{\bf{I}} + \beta_t^2 {{\bf{AA}}^T})$ and  $\{\beta_t\}_{t=1}^{T}$ are a sequence of noise scales satisfying $\beta_{\max} = \beta_1 > \beta_2 > \cdots > \beta_T = \beta_{\min}>0$ \citep{song2020score}. From (\ref{quantized-model-x-tilde}),  an approximation $\tilde{p}({\bf{y}}| {\bf{z}}_t ={\bf{A}} {\bf{x}}_t)$ (called \textit{pseudo-likelihood}) of $p({\bf{y}}| {{\bf{x}}_t})$ can be obtained as \citep{meng2022quantized}  
\begin{align}
    &p({\bf{y}}| {\bf{x}}_t) \simeq \tilde{p}({\bf{y}}| {\bf{z}}_t ={\bf{A}} {\bf{x}}_t) \nonumber \\
    &=   \int \scriptsize{\prod_{m=1}^M  {1}\left( (z_{t,m}+\tilde{n}_{t,m}) \in \mathsf{Q}^{-1}{(y_m)} \right)} \mathcal{N}(\tilde{\bf{n}}_t; {\bf{0}},{\bf{C}}_t^{-1}) d\tilde{\bf{n}}_t,\label{likelihood-integral}
\end{align}
where ${\bf{C}}_t^{-1} = \sigma^2{\bf{I}} + \beta_t^2 {{\bf{AA}}^T}$ and $z_{t,m}, \tilde{n}_{t,m}$ as the $m$-th elements of ${\bf{z}}_t, \tilde{\bf{n}}_t$, respectively, and ${1}(\cdot)$ denotes the indicator function, i.e., it equals  1 if the event in the argument is true and equals 0 otherwise. Furthermore, under the assumption that  $\bf{A}$ is a row-orthogonal matrix such that  ${{\bf{AA}}^T}$ (and thus the covariance matrix ${\bf{C}}_t^{-1}$) becomes diagonal,  \citet{meng2022quantized} obtained a closed-form solution of the \textit{pseudo-likelihood score} $\nabla_{{\bf{x}}_t} \log{\tilde{p}({\bf{y}}| {\bf{z}}_t ={\bf{A}} {\bf{x}}_t)})$, which leads to QCS-SGM. Unfortunately, there is no closed-form solution for general  matrices $\bf{A}$, which is the fundamental limitation of QCS-SGM. For more details of QCS-SGM and SGM, please refer to \citet{meng2022quantized} and \citet{song2019generative,song2020score}, respectively.   

\section{Method}
In this section, to address the inherent limitation of  QCS-SGM \citep{meng2022quantized},  we propose an enhanced variant, termed as QCS-SGM+, for general sensing matrices $\bf{A}$. 

\subsection{A New Perspective} 
Our key insight is that, the  pseudo-likelihood term $\tilde{p}({\bf{y}}| {\bf{z}}_t ={\bf{A}} {\bf{x}}_t)$ (\ref{likelihood-integral})  concerning ${\bf{x}}_t$ can be alternatively interpreted as the partition function (normalization factor) of a posterior distribution concerning $\tilde{\bf{n}}_t$ (rather than ${\bf{x}}_t$), i.e.,
\begin{align}
p(\tilde{\bf{n}}_t|{\bf{y}}) = \frac{f_b(\tilde{\bf{n}}_t) \prod_{m=1}^M f_a(\tilde{{n}}_{t,m}) }{{\tilde{p}({\bf{y}}| {\bf{z}}_t ={\bf{A}} {\bf{x}}_t)}}, \label{eq:post-distribution-EP}
\end{align}
where $f_b(\tilde{\bf{n}}_t) \equiv \mathcal{N}(\tilde{\bf{n}}_t; {\bf{0}},{\bf{C}}_t^{-1}) $  acts as the prior distribution, and $ f_a(\tilde{{n}}_{t,m}) \equiv  {1}\left( (z_{t,m}+\tilde{n}_{t,m}) \in \mathsf{Q}^{-1}{(y_m)} \right)$  acts as the likelihood distribution. As computing the partition function is one fundamental problem in Bayesian inference and various approximate methods have been studied, such a perspective  on $\tilde{p}({\bf{y}}| {\bf{z}}_t ={\bf{A}} {\bf{x}}_t)$ (\ref{likelihood-integral}) as partition function provides us with one solution using well-studied approximate Bayesian inference methods 
\citep{wainwright2008graphical}. 
 
\begin{figure*}[htbp]
\centering
 \includegraphics[width=0.88\textwidth]{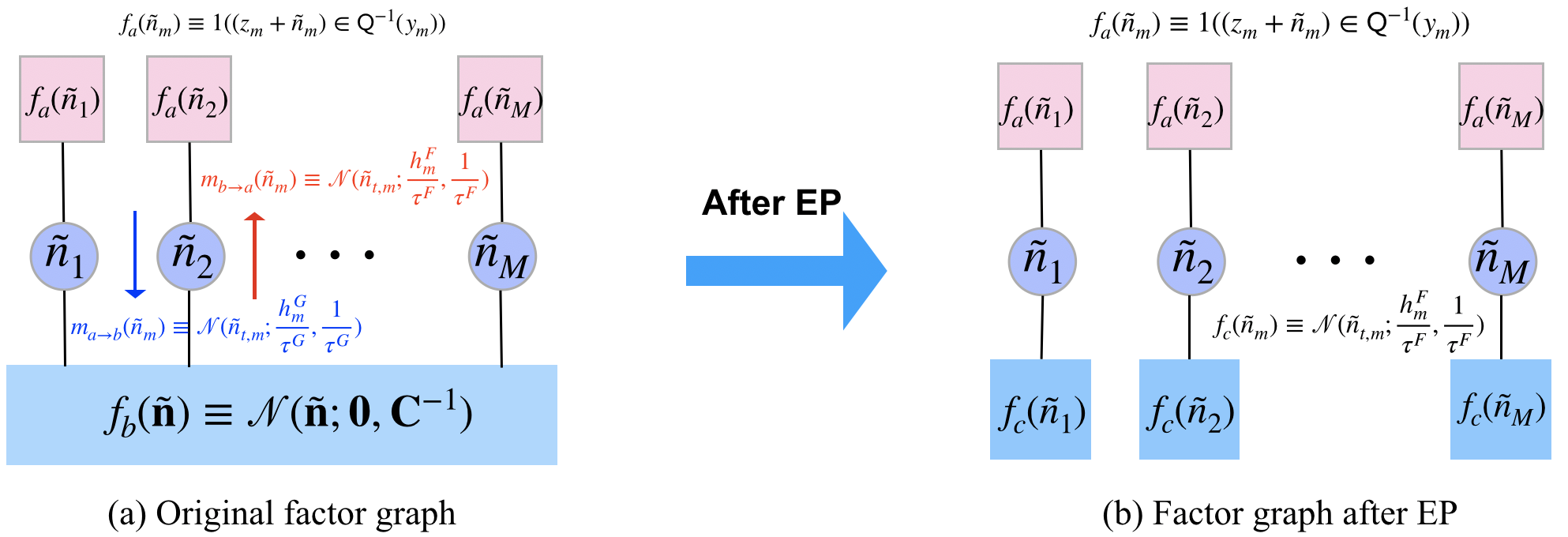}
 \caption{\small{A schematic of the basic idea of QCS-SGM+ in computing the intractable pseudo-likelihood  $\tilde{p}({\bf{y}}| {\bf{z}}_t ={\bf{A}} {\bf{x}}_t)$  for general matrices. The subscript $t$ is dropped for simplicity. We resort to EP to obtain an effective factorized approximation of $f_b({\bf{\tilde{n}}})$} so that a closed-form solution of  $\tilde{p}({\bf{y}}| {\bf{z}}_t ={\bf{A}} {\bf{x}}_t)$ can be achieved, which enables the  computation of otherwise intractable pseudo-likelihood score.} 
 \label{factor_graph}
\end{figure*}

\subsection{Pseudo-Likelihood Score via EP}
Due to its efficacy and  available theoretical guarantee for some general matrices \citep{opper2001tractable,opper2005expectation,takahashi2020semi}, we resort to the well-known expectation propagation (EP) \citep{minka2001expectation} (also known as  moment matching \citep{opper2005expectation}) to approximately compute the intractable partition function $\tilde{p}({\bf{y}}| {\bf{z}}_t={\bf{A}} {\bf{x}}_t)$ in (\ref{eq:post-distribution-EP}), i.e., the pseudo-likelihood  $\tilde{p}({\bf{y}}| {\bf{z}}_t ={\bf{A}} {\bf{x}}_t)$ (\ref{likelihood-integral}). As illustrated in Figure \ref{factor_graph}, the basic idea is to apply EP to derive an effective factorized approximation of $f_b(\tilde{\bf{n}}_t) \equiv \mathcal{N}(\tilde{\bf{n}}_t; {\bf{0}},{\bf{C}}_t^{-1})$. After EP, the original coupled prior node $f_b(\tilde{\bf{n}}_t) $ in Figure \ref{factor_graph} (a) is decoupled into a series of fully-factorized prior nodes $f_c({\tilde{n}_{t,m}}), m=1...M$  in \ref{factor_graph} (b). Hence, a closed-form solution for $\tilde{p}({\bf{y}}| {\bf{z}}_t={\bf{A}} {\bf{x}}_t)$ (\ref{eq:post-distribution-EP}) can be obtained, thereby enabling us to compute the pseudo-likelihood score $\nabla{{\bf{x}}_{t}} \log{\tilde{p}({\bf{y}}| {\bf{z}}_t ={\bf{A}}{\bf{x}}_t)}$. 

The details of our derivation via EP are illustrated as follows. Essentially, it approximates the partition function $\tilde{p}({\bf{y}}| {\bf{z}}_t={\bf{A}} {\bf{x}}_t)$ (\ref{eq:post-distribution-EP}) in three different ways as follows \footnote{The  results might differ in a scaling factor but it does not affect the score function.}:
\begin{align}
  & \tilde{p}({\bf{y}}| {\bf{z}}_t ={\bf{A}} {\bf{x}}_t) \approx \nonumber \\
   & \begin{cases}
      \int {\prod_{m=1}^M}  f_a(\tilde{{n}}_{t,m})  \mathcal{N}(\tilde{n}_{t,m}; \frac{h^F_m}{\tau^F}, \frac{1}{\tau^F}) d\tilde{\bf{n}}_t  & (a)\\
      \int {\prod_{m=1}^M} \mathcal{N}(\tilde{n}_{t,m}; \frac{h^G_m}{\tau^G}, \frac{1}{\tau^G}) f_b(\tilde{\bf{n}}_t)  d\tilde{\bf{n}}_t  & (b)\\
      \int {\prod_{m=1}^M}  \mathcal{N}(\tilde{n}_{t,m}; \frac{h^G_m}{\tau^G}, \frac{1}{\tau^G}) \mathcal{N}(\tilde{n}_{t,m}; \frac{h^F_m}{\tau^F}, \frac{1}{\tau^F}) d\tilde{\bf{n}}_t & (c)
    \end{cases}       
    \label{eq:EP-approx}
\end{align}
Intuitively,  (\ref{eq:EP-approx}-a) approximates the correlated multivariate Gaussian $f_b(\tilde{\bf{n}}_t) \equiv \mathcal{N}(\tilde{\bf{n}}_t; {\bf{0}},{\bf{C}}_t^{-1})$ with a product of independent Gaussians $\prod_{m=1}^M \mathcal{N}(\tilde{n}_{t,m}; \frac{h^F_m}{\tau^F}, \frac{1}{\tau^F})$, (\ref{eq:EP-approx}-b) approximates the non-Gaussian likelihood $f_a(\tilde{{n}}_{t,m}) \equiv 
 {1}\left((z_{t,m}+\tilde{n}_{t,m}) \in \mathsf{Q}^{-1}{(y_m)} \right)$ with $\mathcal{N}(\tilde{n}_{t,m}; \frac{h^G_m}{\tau^G}, \frac{1}{\tau^G})$, and (\ref{eq:EP-approx}-c) combines the two approximations together. In contrast to the original intractable partition function $\tilde{p}({\bf{y}}| {\bf{z}}_t={\bf{A}} {\bf{x}}_t)$ (\ref{eq:post-distribution-EP}), all the three approximations (\ref{eq:EP-approx}-a), (\ref{eq:EP-approx}-b), (\ref{eq:EP-approx}-c) become tractable, leading to three closed-form approximations to the posterior mean $\mathbb{E}[\tilde{n}_{t,m}]$ and variance $\mathbb{V}[\tilde{n}_{t,m}]$ of  $\tilde{n}_{t,m}$ w.r.t. $p(\tilde{\bf{n}}_t|{\bf{y}}) $ (\ref{eq:post-distribution-EP}), which are denoted as $(m_m^a, \chi^a), (m_m^b, \chi^b)$, $(m_m^c, \chi^c)$ respectively, and can be computed as follows 
 \begin{align}
m_m^a & =\frac{h_m^F}{\tau^F} - \frac{2\exp{\left(-\frac{\tilde{u}_{y_m}^2}{2}\right)} - 2\exp{\left(-\frac{\tilde{l}_{y_m}^2}{2}\right)}}{\sqrt{2\pi\tau^F}\Big[\text{erfc}(-\frac{\tilde{u}_{y_m}}{\sqrt{2}}) - \text{erfc}(-\frac{\tilde{l}_{y_m}}{\sqrt{2}})\Big]}, \label{eq:m_ma_result} \\
\chi^a & = \frac{1}{\tau^F} - \frac{1}{M}\sum_{m=1}^{M}\Big[\frac{2\tilde{u}_{y_m}\exp{(-\frac{\tilde{u}_{y_m}^2}{2})} - 2\tilde{l}_{y_m}\exp{(-\frac{\tilde{l}_{y_m}^2}{2})}}{\sqrt{2\pi} \tau^F \Big[\text{erfc}(-\frac{\tilde{u}_{y_m}}{\sqrt{2}}) - \text{erfc}(-\frac{\tilde{l}_{y_m}}{\sqrt{2}})\Big]} \nonumber \\& + \big(m_m^a-\frac{h_m^F}{\tau^F}\big)^2\Big],\label{eq:chi_ma_result}
\end{align}
\begin{align}
m_m^b &= [(\tau^G {\bf{I}} + {\bf{C}}_t)^{-1}\boldsymbol{h}^G]_m, \label{eq:mb} \\
\chi^b &= \text{Tr}[(\tau^G {\bf{I}} + {\bf{C}}_t)^{-1}]/M, \label{eq:chib}\\
m_m^c &=\frac{h_m^G+h_m^F}{\tau^G+\tau^F}, \\
\chi^c &= \frac{1}{\tau^G + \tau^F},
\end{align}
where $\text{erfc}(z)= \frac{2}{\sqrt{\pi}}\int_{z}^{\infty} e^{-{t^2}} dt$ is the complementary error function (erfc) of the standard normal distribution, $\text{Tr}[\cdot]$ is the trace of a matrix, and  $[\cdot]_m$ is the $m$-th element of a vector.
In the special case of 1-bit quantization, i.e., sign measurements as described in (\ref{1-bit model}), the results of $(m_m^a, \chi^a)$  in (\ref{eq:m_ma_result}, \ref{eq:chi_ma_result})  and $g_m$ in (\ref{g-result}) can be further simplified as follows
\begin{align}
m_m^a &=\frac{h_m^F}{\tau^F} + \frac{2y_me^{-\frac{\tilde{l}^2}{2}}}{\sqrt{2\pi\tau^F} \text{erfc}(\frac{y_m\tilde{l}}{\sqrt{2}})},\\
\chi^a &=  \frac{1}{\tau^F}  -\frac{1}{M}\sum_{m=1}^{M}\Big[(m_m^a-\frac{h_m^F}{\tau^F})^2 - \frac{2y_m\tilde{l}e^{-\frac{\tilde{l}^2}{2}}}{\sqrt{2\pi} \tau^F\text{erfc}(\frac{y_m\tilde{l}}{\sqrt{2}})}\Big],\\
g_m &= \frac{y_m\sqrt{2\tau^F}e^{-\frac{\tilde{l}^2}{2}}}{\sqrt{\pi} \text{erfc}(\frac{y_m\tilde{l}}{\sqrt{2}})},
\end{align}
where $\tilde{l}  = - \sqrt{\tau^F} z_{t,m}  - \frac{h_m^F}{\sqrt{\tau^F}}$. 

Subsequently, to determine the associated parameters $(\boldsymbol{h}^F, \tau^F, \boldsymbol{h}^G, \tau^G)$ in  (\ref{eq:EP-approx}), we leverage the moment matching condition of EP \cite{minka2001expectation,opper2005expectation}, leading to the consistency of the associated posterior mean $\mathbb{E}[\tilde{n}_{t,m}]$ and variance $\mathbb{V}[\tilde{n}_{t,m}]$ of  $\tilde{n}_{t,m}$ from all the three approximations in (\ref{eq:EP-approx}), i.e., 
\begin{align}
 m_m^a &= m_m^b = m_m^c, \\
 \chi^a &= \chi^b = \chi^c, 
\end{align}
by which one can easily obtain $(\boldsymbol{h}^F, \tau^F, \boldsymbol{h}^G, \tau^G)$ iteratively. 

Interestingly, the above implementation details of EP can be illustrated via iterative message passing in the corresponding factor graphs \citep{minka2001expectation,wainwright2008graphical}. As shown in Figure \ref{factor_graph}, $\mathcal{N}(\tilde{n}_{t,m}; \frac{h^F_m}{\tau^F}, \frac{1}{\tau^F})$ corresponds to the message $m_{b\to a}(\tilde{n}_m)$ from factor node $f_b$ to factor node $f_a$, while $\mathcal{N}(\tilde{n}_{t,m}; \frac{h^G_m}{\tau^G}, \frac{1}{\tau^G})$ corresponds to the message $m_{a\to b}(\tilde{n}_m)$ from factor node $f_a$ to factor node $f_b$. The three approximations (\ref{eq:EP-approx}-a), (\ref{eq:EP-approx}-b), and (\ref{eq:EP-approx}-c) correspond to the combined results of incoming messages in EP at factor node $f_a$, variable node $\tilde{n}_m$, factor node $f_b$, respectively. 

After EP, as shown in  Figure \ref{factor_graph} (b), we can obtain an alternative factorized representation, leading to a closed-form approximation of $\tilde{p}({\bf{y}}| {\bf{z}}_t ={\bf{A}} {\bf{x}}_t) $ as follows
\begin{align}
\tilde{p}({\bf{y}}| {\bf{z}}_t ={\bf{A}} {\bf{x}}_t) \approx  \frac{e^{ \frac{(h_m^F)^2}{2\tau^F} }}{2}\Big[\text{erfc}(\frac{-\tilde{u}_{y_m}}{\sqrt{2}}) - \text{erfc}(\frac{-\tilde{l}_{y_m}}{\sqrt{2}})\Big], \label{eq:likelihood-result}
\end{align}
where 
\begin{align}
    \tilde{u}_{y_m} &= -\sqrt{\tau^F} z_{t,m} - \frac{h_m^F}{\sqrt{\tau^F}} + u_{y_m}{\sqrt{\tau^F}}, \\
    \tilde{l}_{y_m} &= - \sqrt{\tau^F} z_{t,m}  - \frac{h_m^F}{\sqrt{\tau^F}} + l_{y_m}{\sqrt{\tau^F}}. 
\end{align}
Therefore, from (\ref{likelihood-integral}) and (\ref{eq:likelihood-result}), the intractable pseudo-likelihood score $\nabla_{{\bf{x}}_t} \log p({\bf{y}} \mid {\bf{x}}_t)$ under quantized measurements ${\bf{y}}$ in (\ref{quantized model}) can be approximated as
\begin{align}
    \nabla_{{\bf{x}}_t} \log p({\bf{y}} \mid {\bf{x}}_t) \simeq {\bf{A}}^T {\bf{G}}({\beta_t},{\bf{y}},{\bf{A}},{\bf{z}}_t, \boldsymbol{h}^F, \tau^F), \label{eq:pseudo_grad}
\end{align}
where ${\bf{G}}({\beta_t},{\bf{y}},{\bf{A}},{\bf{z}}_t, \boldsymbol{h}^F, \tau^F) = [g_1,g_2,...,g_M]^T \in \mathbb{R}^{M\times 1}$ with the $m$-th element being 
\begin{align}
 g_m = -\frac{\sqrt{2\tau^F}\Big[\exp{\left(-\frac{\tilde{u}_{y_m}^2}{2}\right)} - \exp{\left(-\frac{\tilde{l}_{y_m}^2}{2}\right)}\Big]}{\sqrt{\pi}\Big[\text{erfc}(-\frac{\tilde{u}_{y_m}}{\sqrt{2}}) - \text{erfc}(-\frac{\tilde{l}_{y_m}}{\sqrt{2}})\Big]}.\label{g-result}
\end{align}

\subsection{Resultant Algorithm: QCS-SGM+}
By combining the pseudo-likelihood score (\ref{eq:pseudo_grad}) approximated via EP and the prior score from SGM, we readily obtain an enhanced version of QCS-SGM, dubbed as QCS-SGM+, using the annealed Langevin dynamics (ALD) \citep{song2019generative}. The details of QCS-SGM+ are shown in Algorithm \ref{posterior-sampling-algorithm} where the number of iterations of EP is denoted as $IterEP$. A scaling factor $\gamma$ is introduced in QCS-SGM+, which is empirically found helpful to further improve the performance. 
\begin{algorithm}[t!]
\caption{QCS-SGM+}
 \label{posterior-sampling-algorithm}
  \KwInput{$\{\beta_t\}_{t=1}^T$, $\epsilon$, $\gamma$, $IterEP$, $K$, $\bf{y,A}$, $\sigma^2$, quantization thresholds $\{[l_q,u_q)|q\in \mathcal{Q}\}$}
  \KwInitialize{${\bf{x}}_1^0\sim \mathcal{U}\left(0,1\right)$}
  \For{$t=1$ {\bfseries to} $T$}{
    $\alpha_t \leftarrow \epsilon \beta_t^2/\beta_T^2$
    
    \For{$k=1$ {\bfseries to} $K$}{
        Draw ${\bf{z}}_t^k \sim \mathcal{N}(\bf{0}, \bf{I}) $
        
        \small{\KwInitialize{$\boldsymbol{h}^F, \tau^F, \boldsymbol{h}^G, \tau^G$}}        
        \For{$it=1$ {\bfseries to} $IterEP$}{  
        $\boldsymbol{h}^G =\frac{\boldsymbol{m}^a}{\chi^a} -\boldsymbol{h}^F$
        
        $\tau^G =\frac{1}{\chi^a} - \tau^F$
        
$\boldsymbol{h}^F = \frac{\boldsymbol{m}^b}{\chi^b} - \boldsymbol{h}^G$

$\tau^F = \frac{1}{\chi^b} -\tau^G$
        }
        {Compute $\nabla_{{\bf{x}}_t} \log p({\bf{y}} \mid {\bf{x}}_t)$ as (\ref{eq:pseudo_grad})}
    
        \footnotesize{${\bf{x}}_{t}^{k} = {\bf{x}}_{t}^{k-1} + \alpha_t \Big[ {\rm{s}}_{\boldsymbol{\theta}}({\bf{x}}_{t}^{k-1},\beta_t) + \gamma \nabla_{{\bf{x}}_t} \log p({\bf{y}} \mid {\bf{x}}_t)\Big] +\sqrt{2\alpha_t}{\bf{z}}_t^k$}
    }
    ${\bf{x}}_{t+1}^{0} \leftarrow {\bf{x}}_{t}^{K}$
   }
\KwOutput{${\bf{\hat{x}}} = {\bf{x}}_{T}^{K}$} 
\end{algorithm}

Note that while it seems from (\ref{eq:mb}, \ref{eq:chib}) that matrix inversion $(\tau^G {\bf{I}} + {\bf{C}}_t)^{-1}$ is needed in each iteration for every ${\bf{C}}_t$,  this is actually not the case since there exists one efficient implementation method using singular value decomposition (SVD) similar to \citet{rangan2019vector,meng2022diffusion}. Specifically, denote  ${\bf{A} = U\Sigma V}^T$ as the SVD of $\bf{A}$ and ${{\Sigma}}^2$ as the element-wise square of singular values, i.e., diagonal elements of $\bf{\Sigma}$, then after some simple algebra, it can be shown that the terms $\boldsymbol{m}^b, \chi^b$ involving a matrix inverse can be efficiently computed as follows
\begin{align}
\boldsymbol{m}^b &= {{\bf{U}} \text{diag}\Big(\frac{\sigma^2 + \beta_t^2 \Sigma^2}{\tau_G (\sigma^2 + \beta_t^2 \Sigma^2) + 1}\Big){\bf{U}}^T}\boldsymbol{h}^G, \label{eq:mb_new} \\ 
\chi^b &= <\frac{\sigma^2 + \beta_t^2 \Sigma^2}{\tau_G (\sigma^2 + \beta_t^2 \Sigma^2) + 1}>, \label{eq:chib_new}
\end{align}
where $<\cdot>$ is the average value of the elements in a vector. It can be seen from (\ref{eq:mb_new}, \ref{eq:chib_new}) that one simply needs to replace the values of $\beta_t, \tau_G$ for different iterations. Hence,  the main computational burden lies in the SVD of sensing matrix $\bf{A}$, but only one time is required in the whole QCS-SGM+.
\begin{figure*}[htbp]
\centering
\subfigure[MNIST, $M=400, \sigma=0.05, \kappa = 10^3$]{
    \begin{minipage}[b]{0.45\textwidth}
    \includegraphics[width=\textwidth]{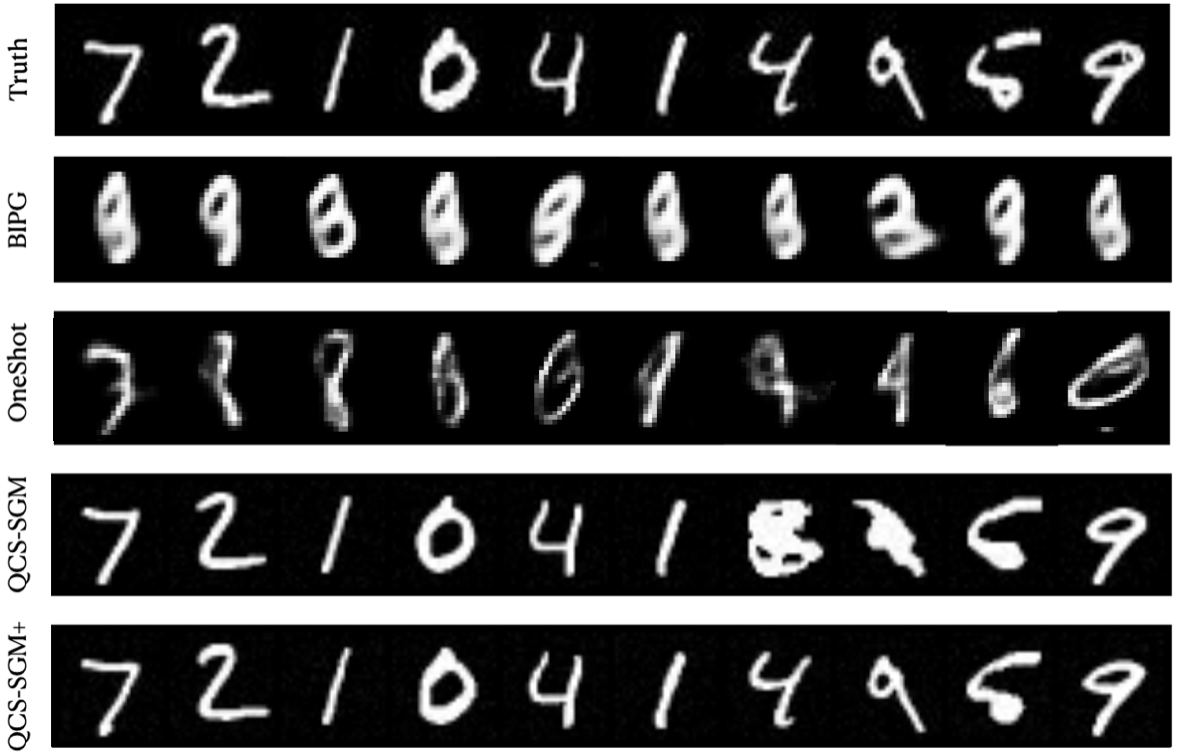}
    \end{minipage}
}
\subfigure[CelebA, $M=4000, \sigma=0.001, \kappa = 10^6$]{
  \begin{minipage}[b]{0.45\textwidth}
    \includegraphics[width=\textwidth]{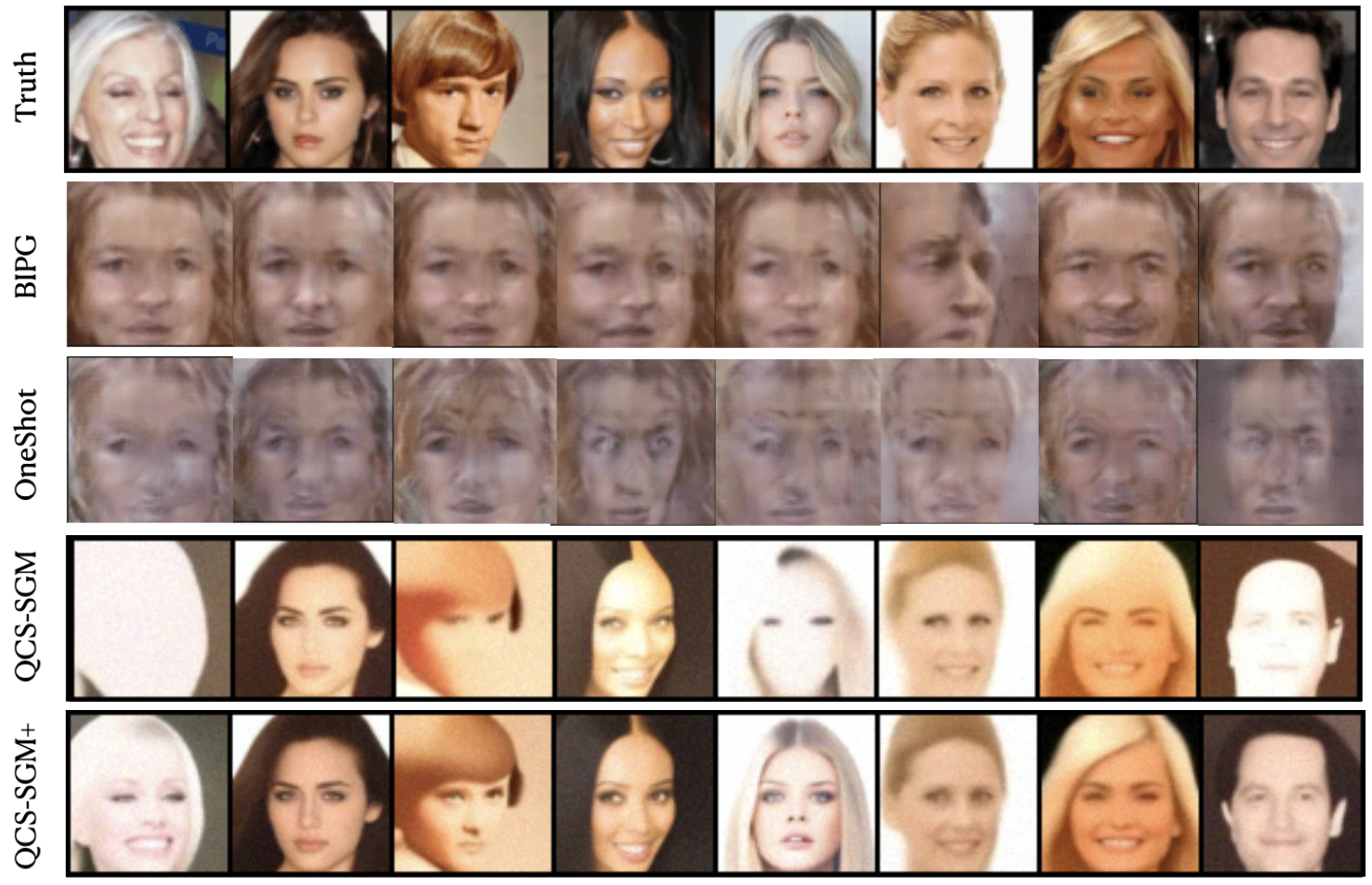}
  \end{minipage}
}
\caption{\small{Qualitative comparisons of different methods under 1-bit CS on MNIST and CelebA for ill-conditioned  $\bf{A}$ ($\kappa=10^3$ for MNIST and $\kappa=10^6$ for CelebA)  when $M< N$. It can be seen that the proposed QCS-SGM+ achieves consistently better
results than other methods. }}
\label{mnist-celeba-compare}
\end{figure*}

\begin{figure*}[!h]
\centering
\subfigure[QCS-SGM]{
    \begin{minipage}[b]{0.45\textwidth}
    \includegraphics[width=\textwidth]{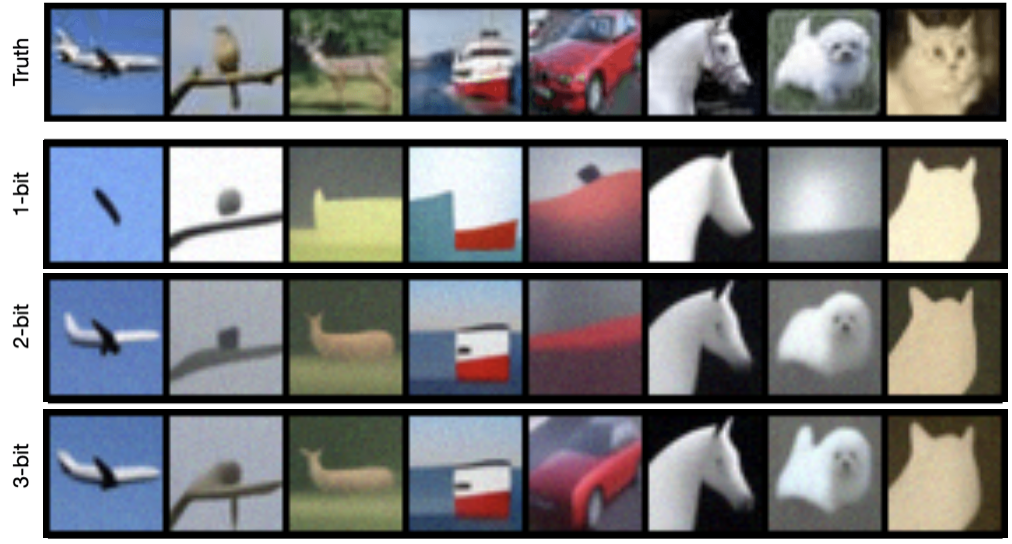}
    \end{minipage}
}
\subfigure[QCS-SGM+]{
  \begin{minipage}[b]{0.45\textwidth}
    \includegraphics[width=\textwidth]{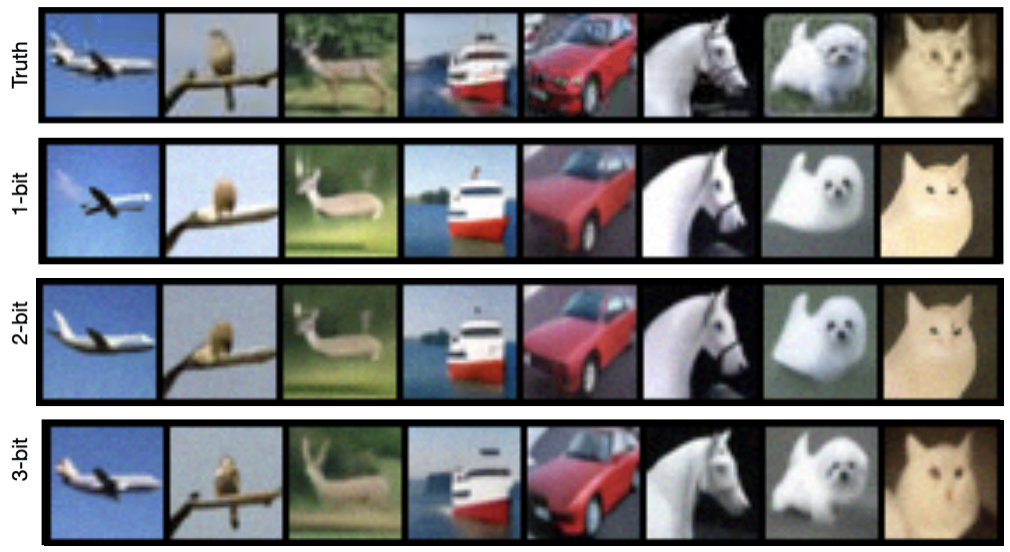}
  \end{minipage}
}
\caption{\small{Reconstructed images on CIFAR-10 with QCS-SGM and QCS-SGM+, respectively, under 1-3 bit CS when the condition number of $\bf{A}$ is 1000, $M=2000, \sigma = 0.1$. It can be seen that the proposed QCS-SGM+ achieves consistently better
results than QCS-SGM.}}
\label{cifar10-3bit}
\end{figure*}

\subsection{Relation to QCS-SGM}
\label{sec:relation}
For row-orthogonal matrices $\bf{A}$, the covariance matrix ${\bf{C}}^{-1}_t = \sigma^2{\bf{I}} + \beta_t^2 {{\bf{AA}}^T}$ becomes diagonal and thus the prior node $f_b(\tilde{\bf{n}}_t) \equiv \mathcal{N}(\tilde{\bf{n}}_t; {\bf{0}},{\bf{C}}_t^{-1})$  in Figure \ref{factor_graph} (a) already fully factorizes and thus  QCS-SGM+ ($\gamma=1$) reduces to QCS-SGM \citep{meng2022quantized}.
For general matrices $\bf{A}$, however, there is no such equivalence even though QCS-SGM can still be applied pretending ${\bf{C}}^{-1}_t$ to be  diagonal, i.e., QCS-SGM directly diagonalizes ${\bf{C}}^{-1}_t$ by extracting its main diagonal elements. The fundamental difference between QCS-SGM and QCS-SGM+ can be better illustrated from the new perspective on ${\tilde{p}({\bf{y}}| {\bf{z}}_t ={\bf{A}} {\bf{x}}_t)}$ as partition function ${\tilde{p}({\bf{y}}| {\bf{z}}_t ={\bf{A}} {\bf{x}}_t)} = \int f_b(\tilde{\bf{n}}_t) \prod_{m=1}^M f_a(\tilde{{n}}_{t,m}) d {\bf{n}}_t$ (\ref{eq:post-distribution-EP}). QCS-SGM naively diagonalizes the correlated Gaussian $f_b(\tilde{\bf{n}}_t) \equiv \mathcal{N}(\tilde{\bf{n}}_t; {\bf{0}},{\bf{C}}_t^{-1})$, which ignores the effect of $\prod_{m=1}^M f_a(\tilde{{n}}_{t,m}) $; by contrast, QCS-SGM+ considers $f_b(\tilde{\bf{n}}_t) \prod_{m=1}^M f_a(\tilde{{n}}_{t,m})$ as a whole and diagonalizes $f_b(\tilde{\bf{n}}_t)$ by explicitly  takes into account the effect of $\prod_{m=1}^M f_a(\tilde{{n}}_{t,m})$, thereby leading to better performances than QCS-SGM for general matrices $\bf{A}$. 

\section{Experiments}
\label{sec:experiements}
We empirically demonstrate the efficacy of the proposed QCS-SGM+ in various scenarios. The source code is available at \textit{https://github.com/mengxiangming/QCS-SGM-plus}. More results, including the detailed hyper-parameter setting,  can be found in the appendix. 

Specifically, we investigate two popular general sensing matrices $\bf{A}$ beyond row-orthogonal:

(a) {\textbf{ill-conditioned matrices}} \citep{rangan2019vector,schniter2016vector,venkataramanan2022estimation,fan2022approximate}: $\bf{A} = V \Sigma U^T$, where $\bf{V}$ and $\bf{U}$ are independent Harr-distributed matrices, and the nonzero singular values of  $\bf{A}$ satisfy $\frac{\lambda_i}{\lambda_{i+1}} = \kappa^{1/M}$, where $\kappa$ is the condition number. Such matrices can have an arbitrary spectral distribution and often arise in practical applications \citep{venkataramanan2022estimation}.  


(b) {\textbf{correlated matrices}} \cite{shiu2000fading,zhu2022unitary}:  ${\mathbf A}$ is constructed as ${\mathbf A}={\mathbf R}_L{\mathbf H}{\mathbf R}_{R}$, where ${\mathbf R}_L={\mathbf R}_1^{\frac{1}{2}}\in{\mathbb R}^{M\times M}$ and ${\mathbf R}_R={\mathbf R}_2^{\frac{1}{2}}\in{\mathbb R}^{N\times N}$, the $(i,j)$th element of both ${\mathbf R}_1$ and ${\mathbf R}_2$ is $\rho^{|i-j|}$ and $\rho$ is termed as the correlation coefficient here, ${\mathbf H}\in{\mathbb R}^{M\times N}$ is a random matrix whose elements are drawn i.i.d. from ${\mathcal N}(0,1)$.

\textbf{Datasets}: We consider 
three  popular datasets: MNIST \citep{mnist_dataset} , CIFAR-10 \citep{krizhevsky2009learning}, and CelebA  \citep{liu2015faceattributes}. For CelebA dataset, we cropped each face image to a $64\times 64$ RGB image, resulting in $N=64\times 64\times3=12288$ inputs per image. 

\textbf{QCS-SGM+}: Same as QCS-SGM \citep{meng2022quantized}, we adopt the  NCSNv2 \citep{song2020improved} in all cases. For MNIST, the NCSNv2 was trained  with a similar training set up as CIFAR-10 in \citet{song2020improved}, while for CIFAR-10, and CelebA, we use the pre-trained models from \url{https://drive.google.com/drive/folders/1217uhIvLg9ZrYNKOR3XTRFSurt4miQrd}. 

\textbf{Comparison Methods}: We compare QCS-SGM+ with not only the state-of-the-art QCS-SGM \citep{meng2022quantized}, but also two other leading methods before QCS-SGM, namely BIPG \citep{liu2020sample} and  OneShot \citep{liu2022non}.

\renewcommand{\dblfloatpagefraction}{1}
\begin{figure*}[t]
\centering
 \includegraphics[width=0.88\textwidth]{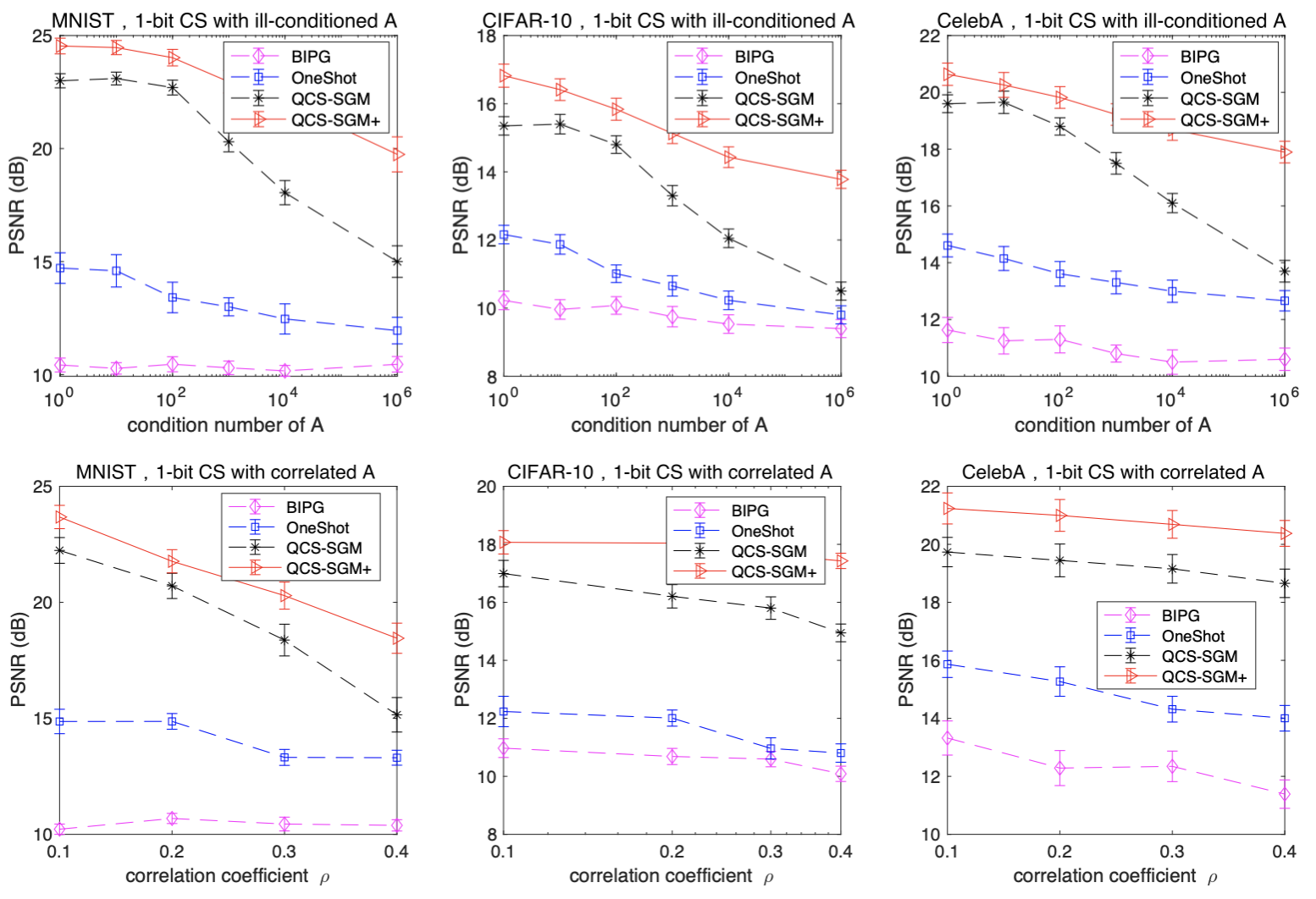}
 \caption{\small{Quantitative comparisons between QCS-SGM+ and existing methods under 1-bit CS for MNIST, CIFAR-10, and CelebA when the sensing matrices $\bf{A}$ are ill-conditioned and correlated, respectively. The number of measurements are $M=400,2000,4000$ for MNIST, CIFAR-10, and CelebA, respectively, all satisfying $M<N$. Standard error bars for the results are also shown.}}
 \label{metrics_compare_mnist_celeba}
\end{figure*}

\renewcommand{\dblfloatpagefraction}{1}
\begin{figure*}[!h]
\centering
 \includegraphics[width=1\textwidth]{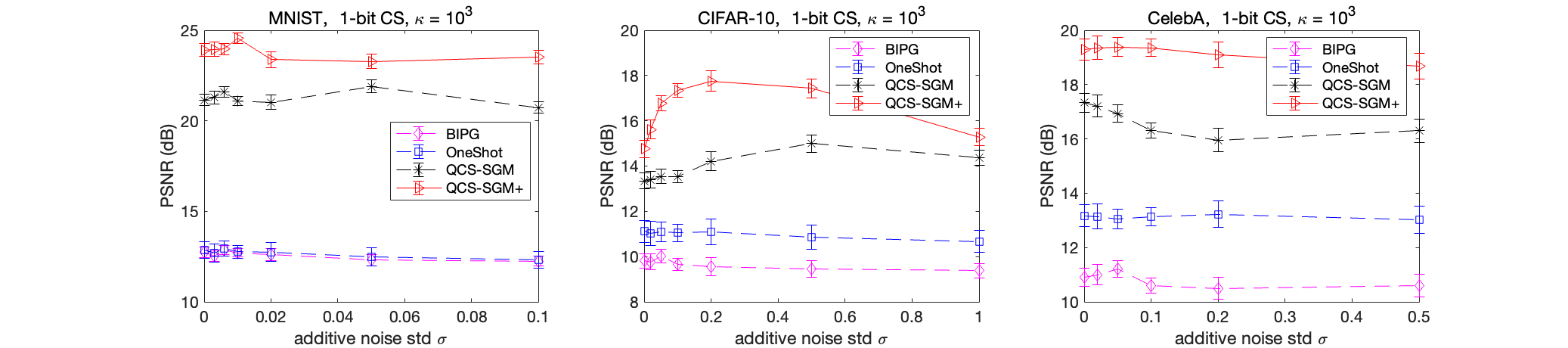}
 \caption{\small{Quantitative comparisons between QCS-SGM+ and other methods under 1-bit CS with different levels of Gaussian noise for ill-conditioned $\bf{A}$ with condition number $\kappa=10^3$. The number of measurements are $M=400,2000,4000$ for MNIST, CIFAR-10, and CelebA, respectively, all satisfying $M< N$. Standard error bars for the results are also shown.}}
 \label{noise-effect}
\end{figure*}

\subsection{Qualitative Results}
Figure \ref{mnist-celeba-compare} shows some qualitative results of QCS-SGM+ and QCS-SGM, BIFG, and OneShot under 1-bit CS for ill-conditioned $\bf{A}$. It can be seen from Figure \ref{mnist-celeba-compare} that the proposed QCS-SGM+ achieves remarkably better performances than all previous methods and can well reconstructs the target images from only a few $M< N $ sign measurements even when the sensing matrix $\bf{A}$ is highly ill-conditioned. To further demonstrate the efficacy of QCS-SGM+ under different   quantization resolutions (e.g., 1-3 bits), Figure \ref{cifar10-3bit} shows some results of QCS-SGM and QCS-SGM+, respectively, with condition number of $\bf{A}$ being $\kappa=10^3$ and $\sigma = 0.1$. It can be seen that QCS-SGM+ is able to generate clear images even with $M< N$ 1-3 bit noisy quantized measurements, whereas the original QCS-SGM yields only vague or blurry results.

\subsection{Quantitative Results}
The quantitative comparison in terms of the peak signal-to-noise ratio (PSNR) is evaluated.  First, Figure \ref{metrics_compare_mnist_celeba} illustrates the PSNR results of QCS-SGM+ and QCS-SGM, in the case of 1-bit CS with MNIST, CIFAR-10, and CelebA, for ill-conditioned $\bf{A}$ and correlated $\bf{A}$, respectively. It can be seen that in all cases, the proposed QCS-SGM+ achieves much better performances, demonstrating the superiority of QCS-SGM+ over QCS-SGM for more general sensing matrices $\bf{A}$. We also evaluate the effect of Gaussian noise ${\bf{n}}\sim \mathcal{N}({\bf{n}};0,\sigma^2 {\bf{I}})$ by conducting experiments on 1-bit CS with different levels of noise standard deviation (std) $\sigma$. As shown in Figure \ref{noise-effect}, due to the potential dithering effect, QCS-SGM+ with noise can sometimes achieve slightly better results than that without noise. Generally, QCS-SGM+ is  robust to noise and can achieve good results in a large range of $\bf{n}$, while significantly outperforming QCS-SGM.

\section{Conclusion}
\label{sec:Conclusion}
In this paper, we  propose an improved version of QCS-SGM, termed as QCS-SGM+, for quantized compressed sensing under general sensing matrices. By viewing the likelihood computation as a Bayesian inference problem, QCS-SGM+  approximates the intractable likelihood score using the well-known EP.  To verify the effectiveness of QCS-SGM+, we conducted experiments on a variety of baseline datasets, demonstrating that QCS-SGM+ significantly outperforms QCS-SGM by a large margin for general sensing matrices. There are several limitations of QCS-SGM+. For example, QCS-SGM+ requires EP message passing, which is computationally slower than QCS-SGM. Also, same as QCS-SGM, it requires a large number of iterations to generate one posterior sample. As future work, it is important to further reduce the complexity of QCS-SGM+ and develop more efficient alternatives with more advanced diffusion models. Moreover, a theoretical analysis of both QCS-SGM and QCS-SGM+ is also an important future direction.

\section*{Acknowledgements}
This work was supported by NSFC No. 62306277,  China, and JSPS KAKENHI Nos. 17H00764, 18K11463, and 19H01812, 22H05117, and JST
CREST Grant Number JPMJCR1912, Japan.

\bibliography{aaai24.bib}

\clearpage
\onecolumn
\appendix

\section{Hyper-parameter Setting of QCS-SGM+}
\label{appendx-hyper-parameter}
As shown in Algorithm \ref{posterior-sampling-algorithm}, there are several hyper-parameters in the proposed QCS-SGM+, including 
$\{\beta_t\}_{t=1}^T$, $\epsilon$, $\gamma$, $IterEP$, $K$. Among them, the values of $\{\beta_t\}_{t=1}^T$ and $K$ are exactly the same as the pre-trained SGM model, i.e., once the pre-trained SGM model is given, $\{\beta_t\}_{t=1}^T$ and $K$ are fixed as the same value as the SGM. The step size $\epsilon$ is set to be the same as QCS-SGM, which is a constant value (here, same as QCS-SGM, we set $\epsilon$ to be $\epsilon=0.002$ in all the experiments), though some further improvement can be expected with a careful fine-tuning of it. 

\begin{figure*}[htbp]
\centering
 \includegraphics[width=1.0\textwidth]{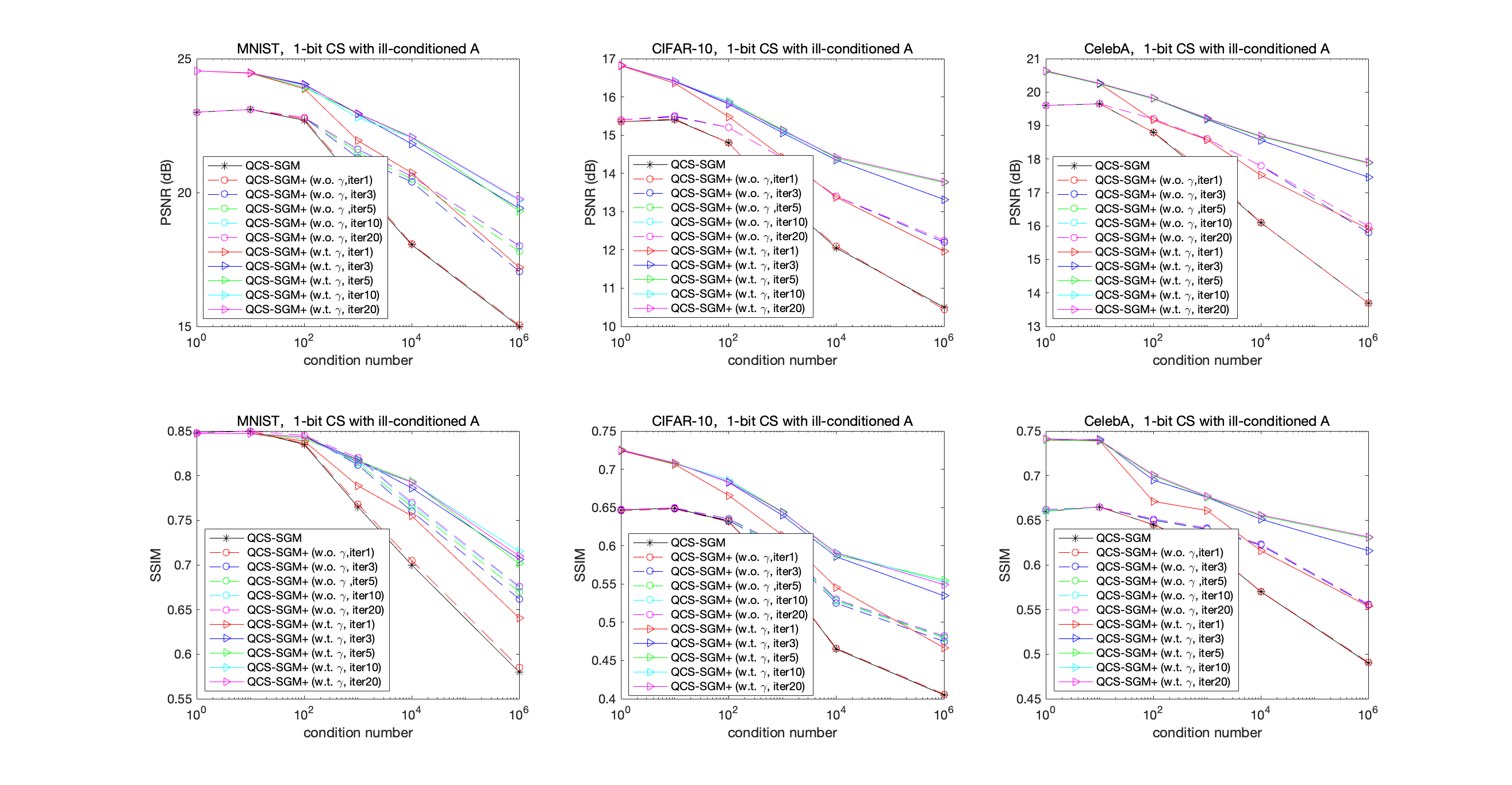}
 \caption{1-bit CS results of QCS-SGM+ on various datasests with (w.t.) and without (w.o.) scaling factor $\gamma$ for different number of EP iterations $IterEP$. In this example, the sensing matrix $\bf{A}$  used is ill-conditioned with different condition numbers. Note that both PSNR and SSIM results are shown.}
 \label{results_iterations}
\end{figure*}

Regarding the scaling factor $\gamma$ of QCS-SGM+, it is set as follows
\begin{align}
\gamma = \xi  \frac{\lVert{{\rm{s}}_{\boldsymbol{\theta}}({\bf{x}}_{t}^{k-1},\beta_t)\rVert}}{\lVert\nabla_{{\bf{x}}_t} \log p_{\beta_t}({\bf{y}} \mid {\bf{x}}_t)\rVert}, \label{eq:scaling-form}
\end{align}
where $\xi>0$ is another induced scalar hyper-parameter. In the original QCS-SGM, no $\gamma$ is introduced, i.e., it is simply set to be a fixed value $\gamma=1$. However, it is empirically found that adding a proper scaling factor $\gamma$ leads to a better performance (as shown in Figure \ref{results_iterations}). The choice of the  form in  (\ref{eq:scaling-form}) is  inspired from \citet{jalal2021robust}. 

Regarding the number of iterations $IterEP$ of QCS-SGM+, it is empirically found that 3-5 iterations are enough in most cases (as shown in Figure \ref{results_iterations}). As a result, if there is no  further illustration, the experimental results shown in this paper are all  using $IterEP=5$ by default.   

To illustrate the effect of $\gamma$ and $IterEP$ of QCS-SGM+, Figure \ref{results_iterations} shows 1-bit CS results of QCS-SGM+ on various datasests with (w.t.) and without (w.o.) scaling factor $\gamma$ for different number of EP iterations $IterEP$. Specifically, we set $\xi = 0.5$ for both MNIST and CIFAR-10, and $\xi = 0.3$ for CelebA for QCS-SGM+ w.t. $\gamma$ in all the related experiments. It can be seen that in both cases with and without $\gamma$, QCS-SGM+ converges with 3-5 EP iterations. Interestingly, when the sensing matrix $\bf{A}$ deviates from row-orthogonal matrix, i.e., the condition number becomes much larger than 1, QCS-SGM+ w.o. $\gamma$ apparently outperforms  the original QCS-SGM, demonstrating the efficacy of EP in QCS-SGM+. Furthermore, QCS-SGM+ w.t. $\gamma$ further improves QCS-SGM+ w.o. $\gamma$ by introducing the scaling factor $\gamma$. Overall, the proposed QCS-SGM+ significantly outperforms the original QCS-SGM.

\section{Additional Results}
All the experiments are conducted on one NVIDIA Tesla V100. In this appendix, we show some additional results apart from those in the main text. 

\label{appendix-varying-noise}

\label{appendix-additional-results}
\subsection{Quantitative Results in terms of SSIM}
\begin{figure*}[t]
\centering
 \includegraphics[width=0.83\textwidth]{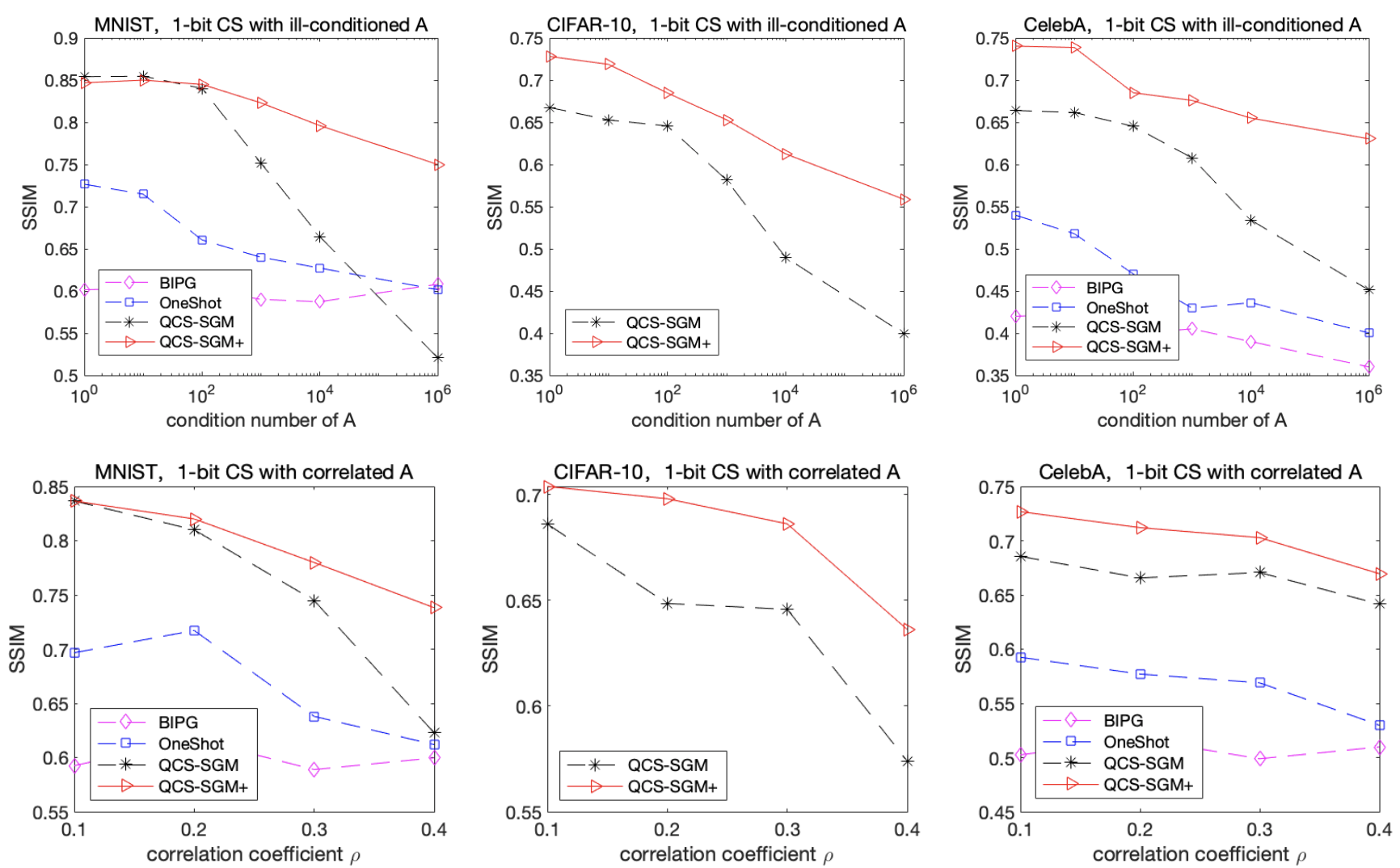}
 \caption{\small{Quantitative comparisons (in terms of SSIM) of the proposed QCS-SGM+ and the original QCS-SGM under 1-bit CS for MNIST, CIFAR-10, and CelebA datasets when the measurement matrices $\bf{A}$ are ill-conditioned and correlated, respectively. The number of measurements are set to be $M=400,2000,4000$ for MNIST, CIFAR-10, and CelebA, respectively, all satisfying $M< N$. 
 In all cases, QCS-SGM+ apparently outperforms QCS-SGM.}}
 \label{metrics_compare_mnist_celeba_ssim}
\end{figure*}

\begin{figure*}[htbp]
\centering
 \includegraphics[width=1\textwidth]{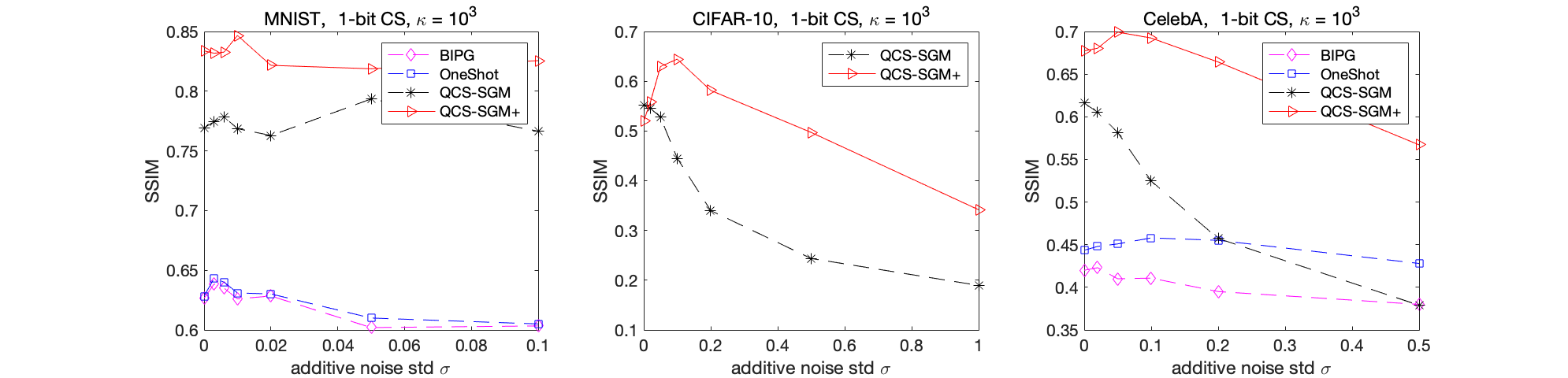}
 \caption{\small{Quantitative comparisons (in terms of SSIM) of the proposed QCS-SGM+ and the original QCS-SGM under 1-bit CS with different levels of additive Gaussian noise for ill-conditioned $\bf{A}$ with condition number $\kappa=10^3$. The number of measurements are set to be $M=400,2000,4000$ for MNIST, CIFAR-10, and CelebA, respectively, all satisfying $M< N$. In all cases, QCS-SGM+ apparently outperforms QCS-SGM.}}
 \label{noise-effect_ssim}
\end{figure*}

In this section, we provide additional quantitative results in terms of  the structural similarity index
measure (SSIM). As shown in Figure \ref{metrics_compare_mnist_celeba_ssim} and Figure \ref{noise-effect_ssim}, similar to the PSNR metric, the proposed QCS-SGM+ significantly outperforms QCS-SGM as well as BIPG and OneShot in terms of SSIM.  Note that results of BIPG and OneShot for CIFAR-10 are not shown since there is a lack of pre-trained model in their open-sourced code. 

\renewcommand{\dblfloatpagefraction}{.9}

\renewcommand{\dblfloatpagefraction}{.9}

\subsection{Additional Qualitative Results}
\begin{figure*}[htbp]
\centering
 \includegraphics[width=1.0\textwidth]{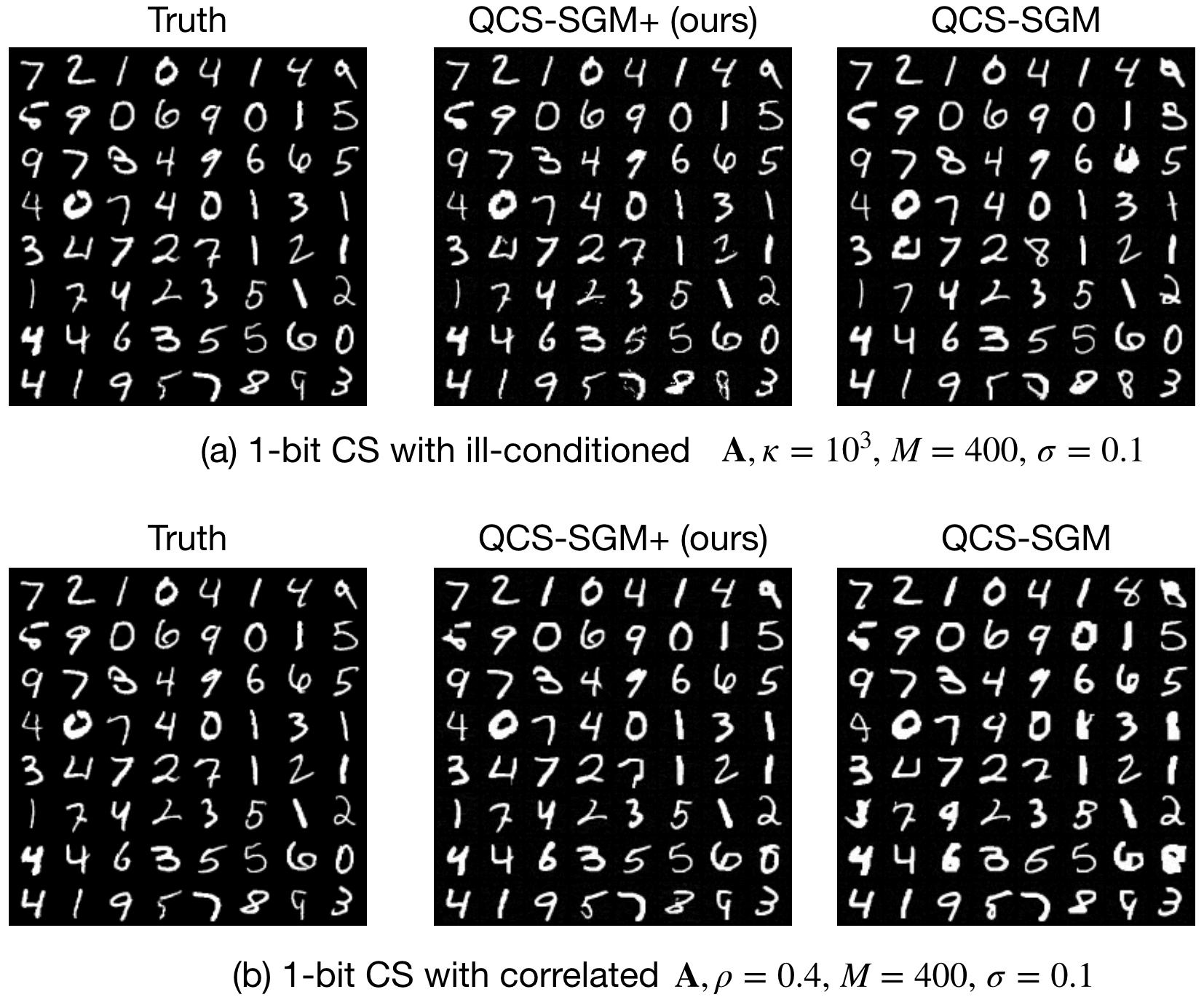}
 \caption{Comparison of QCS-SGM and QCS-SGM+ (ours) for 1-bit CS on MNIST with (a) ill-conditioned $\bf{A}$ and (b) correlated $\bf{A}$, respectively. It can be seen that in both cases QCS-SGM+ outperforms QCS-SGM remarkably.}
 \label{mnist_compare_appendix}
\end{figure*}

\begin{figure*}[htbp]
\centering
\begin{minipage}[t]{0.6\textwidth}
    \centering
    \includegraphics[width=\textwidth]{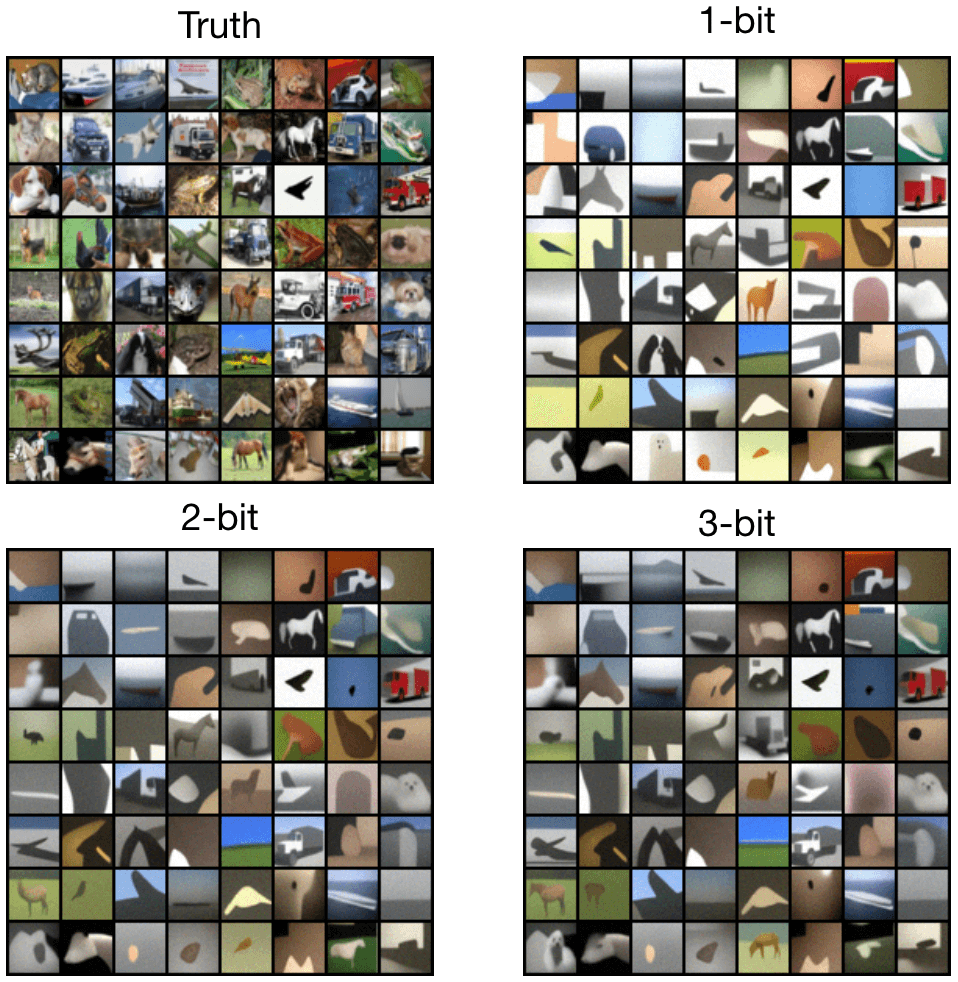}
    \caption*{(a) QCS-SGM}
\end{minipage}
\hfill
\begin{minipage}[t]{0.6\textwidth}
    \centering
    \includegraphics[width=\textwidth]{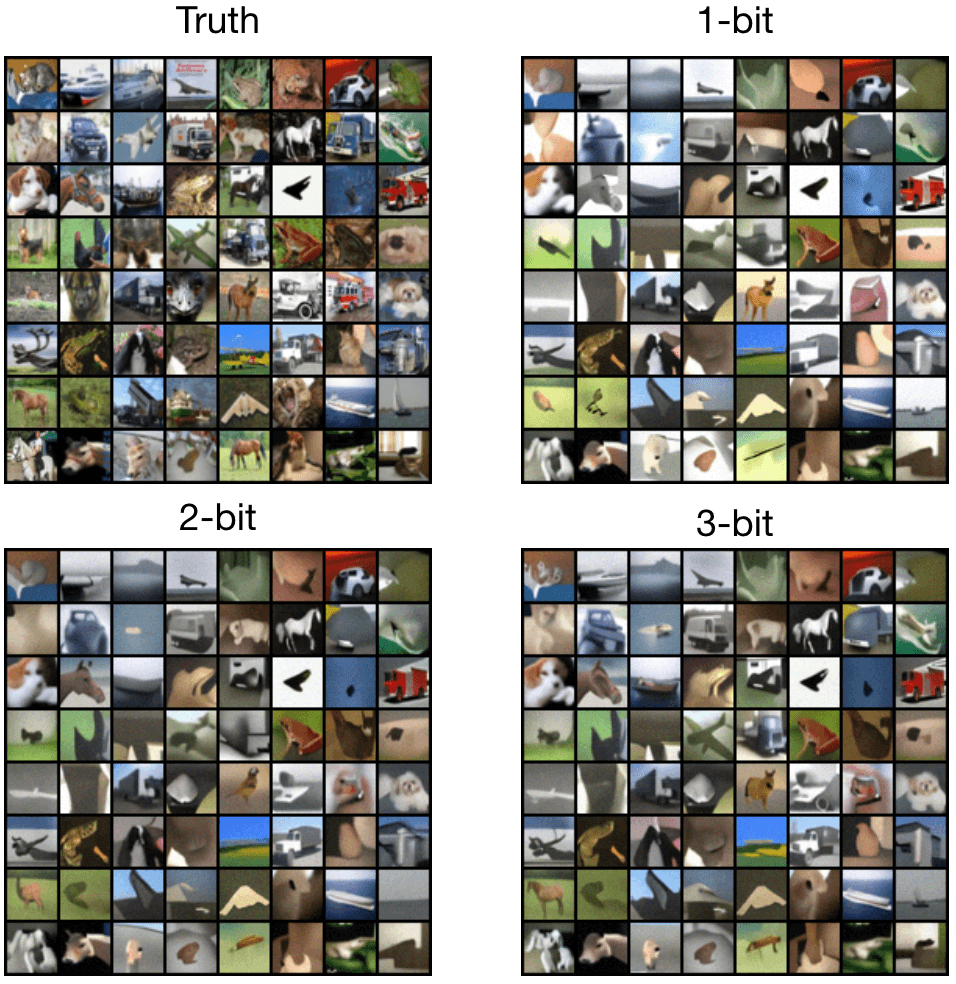}
    \caption*{(b) QCS-SGM+ (ours)}
\end{minipage}
\captionsetup{justification=centering} 
\caption{Results of QCS-SGM+ and QCS-SGM under 1-3 bit CS on CIFAR-10 for ill-conditioned $\mathbf{A}$ ($\kappa=10^3$) when $M=2000, \sigma=0.1$. QCS-SGM+ outperforms QCS-SGM remarkably.}
\label{cifar10-kappa1e3-compare}
\end{figure*}

\begin{figure*}[htbp]
\centering
\begin{minipage}[t]{0.6\textwidth}
    \centering
    \includegraphics[width=\textwidth]{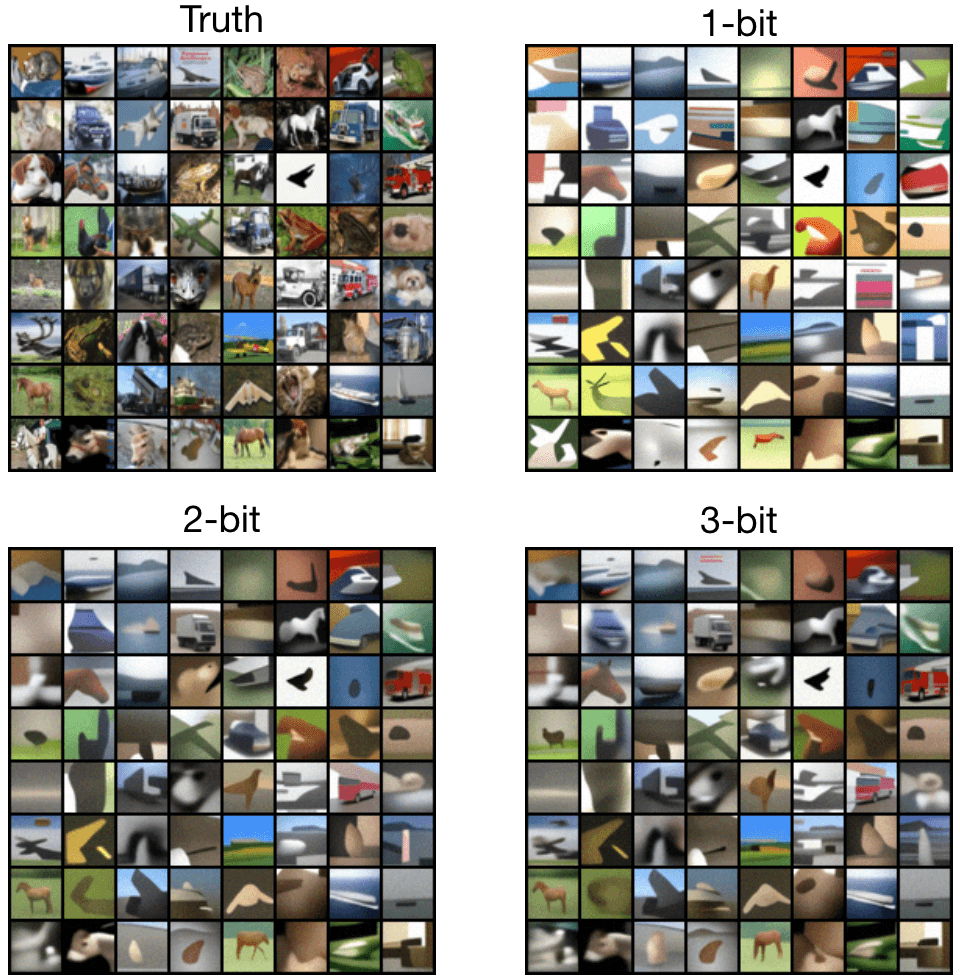}
    \caption*{(a) QCS-SGM}
\end{minipage}
\hfill
\begin{minipage}[t]{0.6\textwidth}
    \centering
    \includegraphics[width=\textwidth]{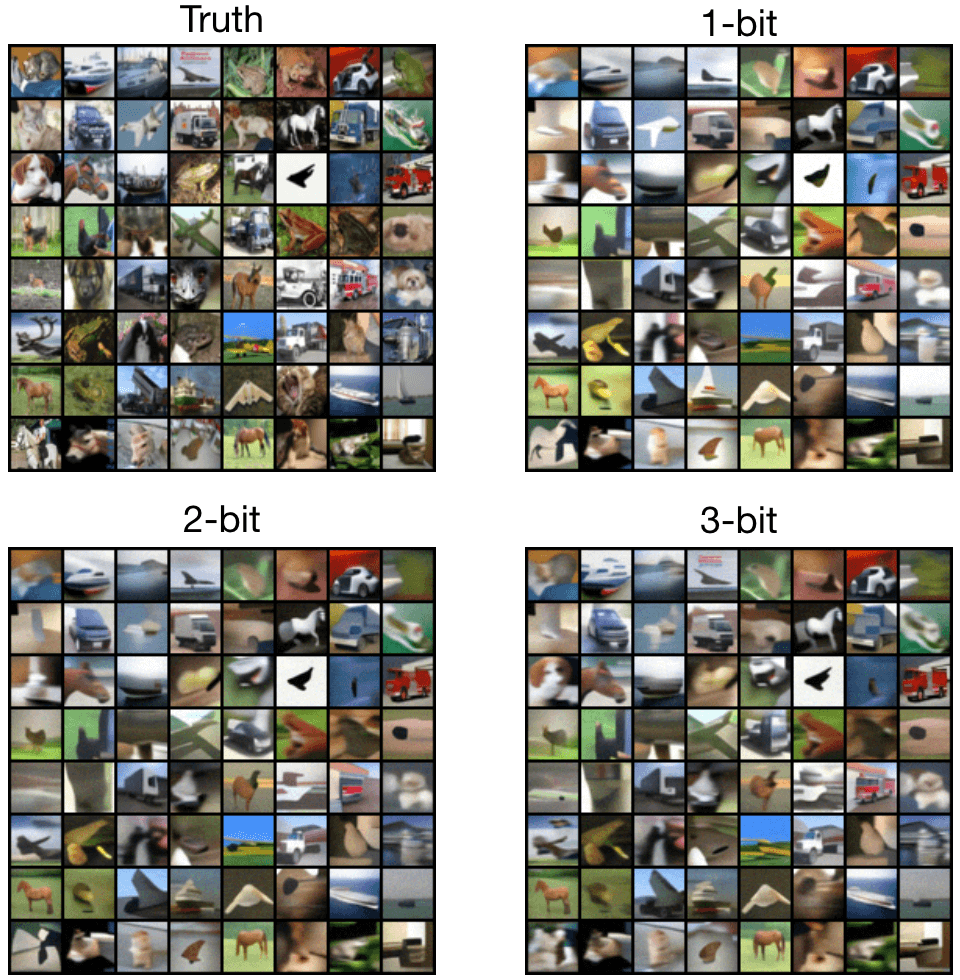}
    \caption*{(b) QCS-SGM+ (ours)}
\end{minipage}
\captionsetup{justification=centering} 
\caption{Results of QCS-SGM+ and QCS-SGM under 1-3 bit CS on CIFAR-10 for correlated $\mathbf{A}$ ($\rho=0.4$) when $M=2000, \sigma=0.1$. QCS-SGM+ outperforms QCS-SGM remarkably.}
\label{cifar10-rho04-compare}
\end{figure*}

\begin{figure*}[htbp]
\centering
\subfigure[Truth]{
    \begin{minipage}[b]{0.45\textwidth}
    \includegraphics[width=\textwidth]{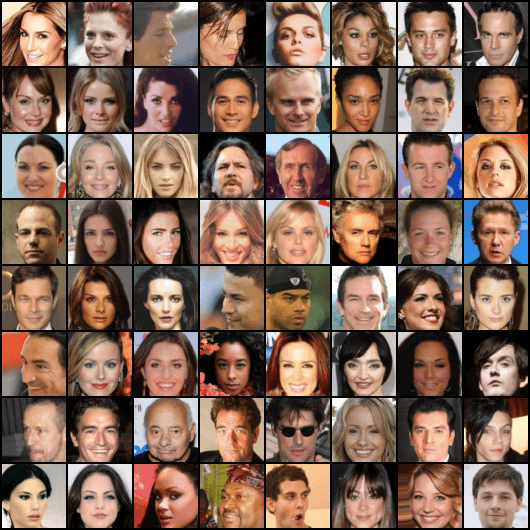}
    \end{minipage}
}
\subfigure[1-bit]{
  \begin{minipage}[b]{0.45\textwidth}
    \includegraphics[width=\textwidth]{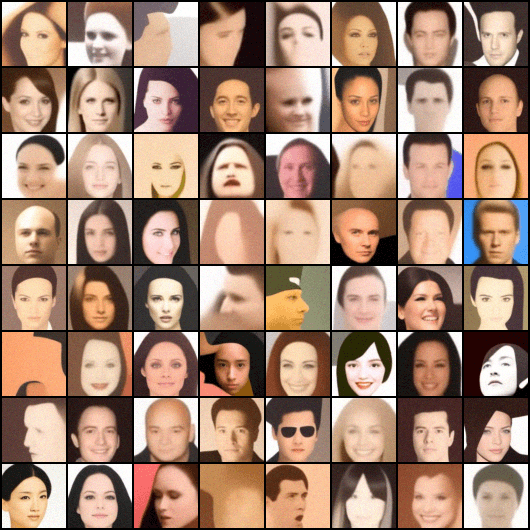}
  \end{minipage}
}
\subfigure[2-bit]{
    \begin{minipage}[b]{0.45\textwidth}
    \includegraphics[width=\textwidth]{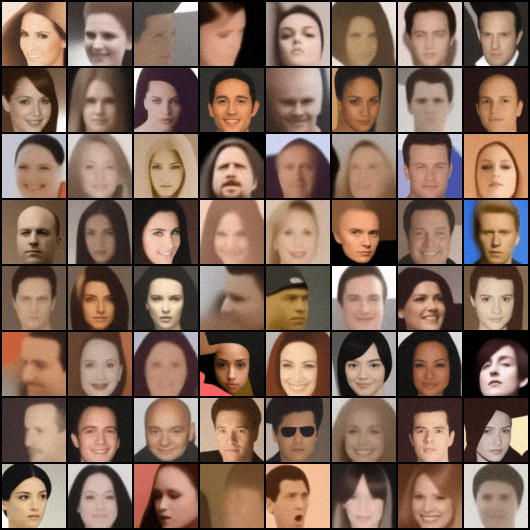}
    \end{minipage}
}
\subfigure[3-bit]{
  \begin{minipage}[b]{0.45\textwidth}
    \includegraphics[width=\textwidth]{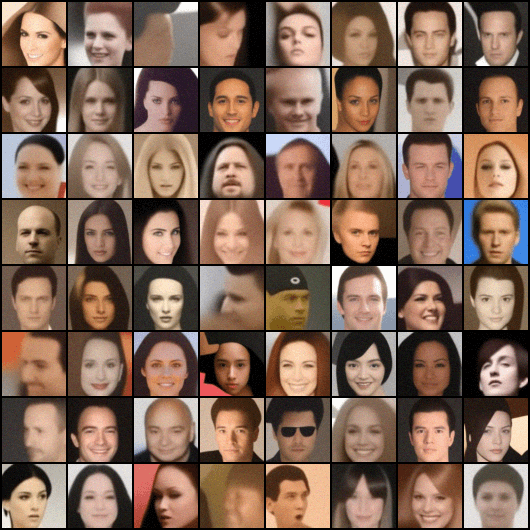}
  \end{minipage}
}
\caption{Results of QCS-SGM under 1-3 bit CS on CelebA for ill-conditioned $\mathbf{A}$ ($\kappa=10^3$) when $M=4000, \sigma=0.1$.  }
\label{celeba-ill-qcs-sgm}
\end{figure*}

\begin{figure*}[htbp]
\centering
\subfigure[Truth]{
    \begin{minipage}[b]{0.45\textwidth}
    \includegraphics[width=\textwidth]{./figures/celeba_true_appendix.png}
    \end{minipage}
}
\subfigure[1-bit]{
  \begin{minipage}[b]{0.45\textwidth}
    \includegraphics[width=\textwidth]{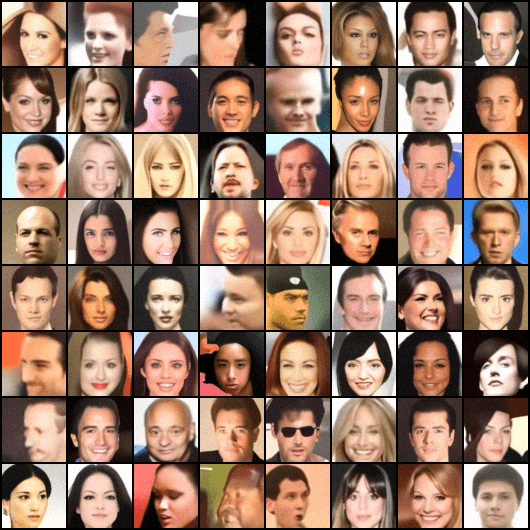}
  \end{minipage}
}
\subfigure[2-bit]{
    \begin{minipage}[b]{0.45\textwidth}
    \includegraphics[width=\textwidth]{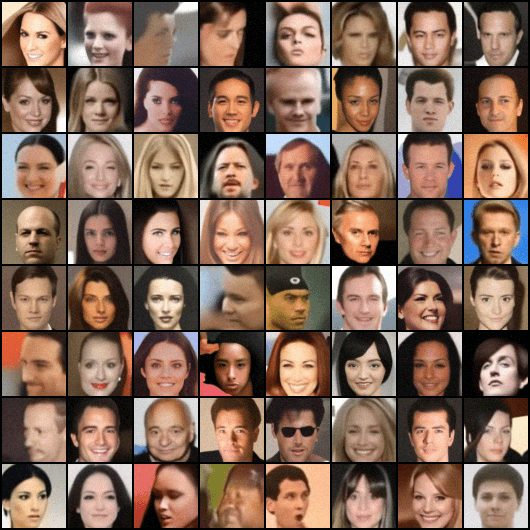}
    \end{minipage}
}
\subfigure[3-bit]{
  \begin{minipage}[b]{0.45\textwidth}
    \includegraphics[width=\textwidth]{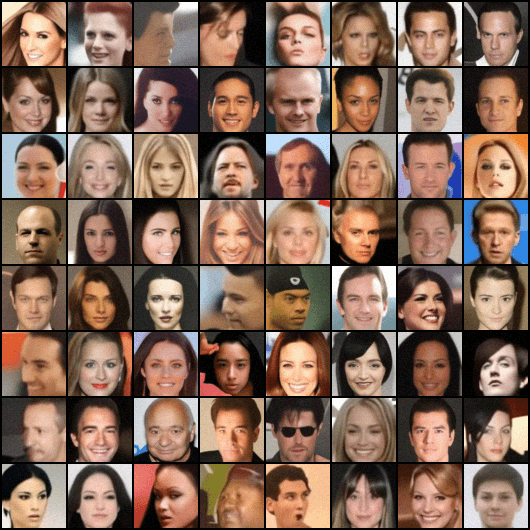}
  \end{minipage}
}
\caption{Results of QCS-SGM+ (ours) under 1-3 bit CS on CelebA for ill-conditioned $\mathbf{A}$ ($\kappa=10^3$) when $M=4000, \sigma=0.1$.  It can be seen that QCS-SGM+ outperforms QCS-SGM in Figure \ref{celeba-ill-qcs-sgm} remarkably. }
\label{celeba-ill-qcs-sgm-plus}
\end{figure*}

\begin{figure*}[htbp]
\centering
\subfigure[Truth]{
    \begin{minipage}[b]{0.45\textwidth}
    \includegraphics[width=\textwidth]{./figures/celeba_true_appendix.png}
    \end{minipage}
}
\subfigure[1-bit]{
  \begin{minipage}[b]{0.45\textwidth}
    \includegraphics[width=\textwidth]{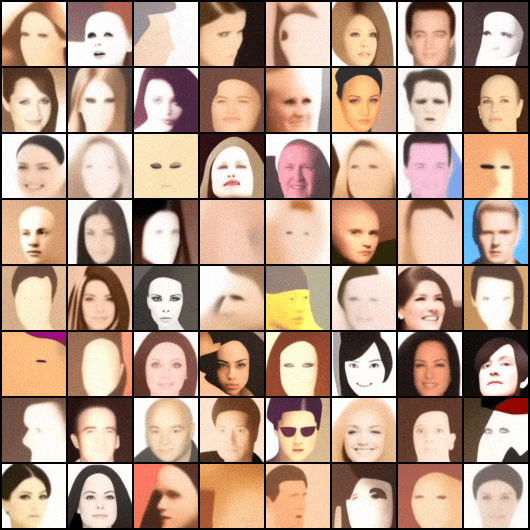}
  \end{minipage}
}
\subfigure[2-bit]{
    \begin{minipage}[b]{0.45\textwidth}
    \includegraphics[width=\textwidth]{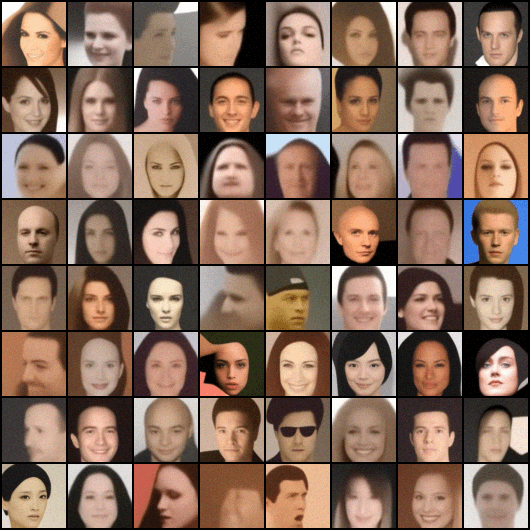}
    \end{minipage}
}
\subfigure[3-bit]{
  \begin{minipage}[b]{0.45\textwidth}
    \includegraphics[width=\textwidth]{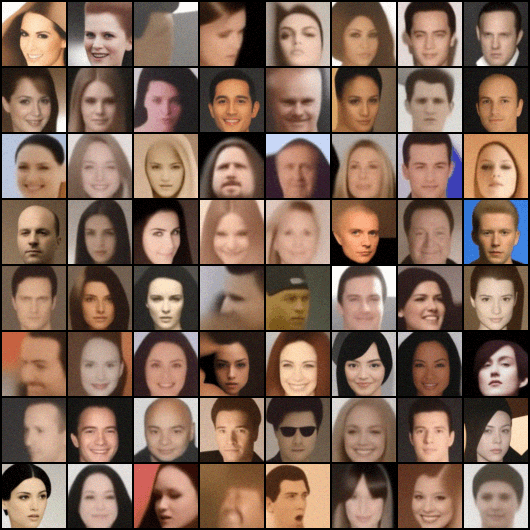}
  \end{minipage}
}
\caption{Results of QCS-SGM under 1-3 bit CS on CelebA for ill-conditioned $\mathbf{A}$ ($\kappa=10^6$) when $M=4000, \sigma=0.1$. }
\label{celeba-ill-qcs-sgm-1e6}
\end{figure*}

\begin{figure*}[htbp]
\centering
\subfigure[Truth]{
    \begin{minipage}[b]{0.45\textwidth}
    \includegraphics[width=\textwidth]{./figures/celeba_true_appendix.png}
    \end{minipage}
}
\subfigure[1-bit]{
  \begin{minipage}[b]{0.45\textwidth}
    \includegraphics[width=\textwidth]{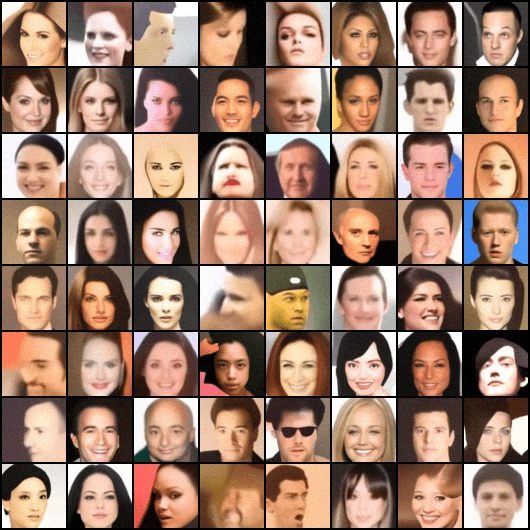}
  \end{minipage}
}
\subfigure[2-bit]{
    \begin{minipage}[b]{0.45\textwidth}
    \includegraphics[width=\textwidth]{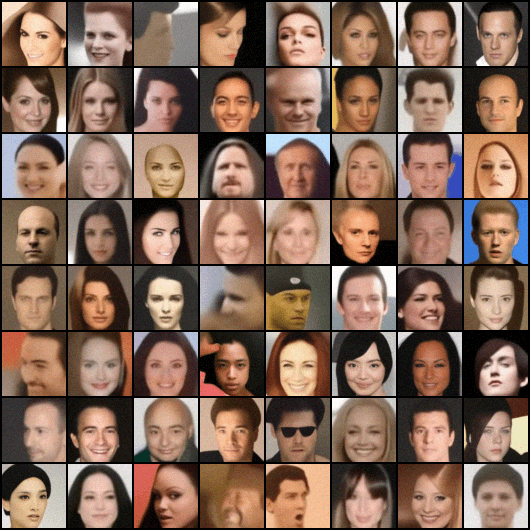}
    \end{minipage}
}
\subfigure[3-bit]{
  \begin{minipage}[b]{0.45\textwidth}
    \includegraphics[width=\textwidth]{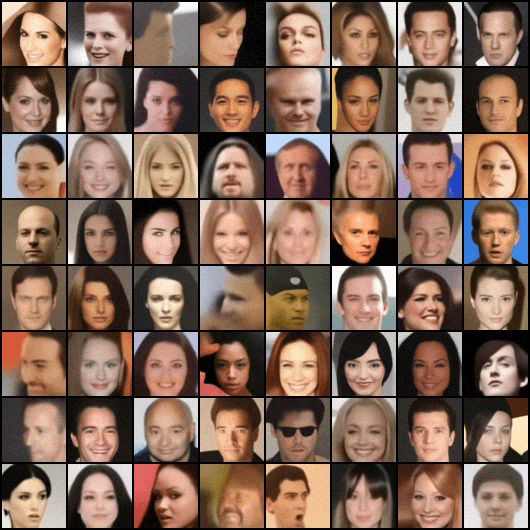}
  \end{minipage}
}
\caption{Results of QCS-SGM+ (ours) under 1-3 bit CS on CelebA for ill-conditioned $\mathbf{A}$ ($\kappa=10^3$) when $M=4000, \sigma=0.1$.  It can be seen that QCS-SGM+ outperforms QCS-SGM in Figure \ref{celeba-ill-qcs-sgm-1e6} remarkably.}
\label{celeba-ill-qcs-sgm-plus-1e6}
\end{figure*}

\begin{figure*}[!h]
\centering
\subfigure[Truth]{
    \begin{minipage}[b]{0.45\textwidth}
    \includegraphics[width=\textwidth]{./figures/celeba_true_appendix.png}
    \end{minipage}
}
\subfigure[1-bit]{
  \begin{minipage}[b]{0.45\textwidth}
    \includegraphics[width=\textwidth]{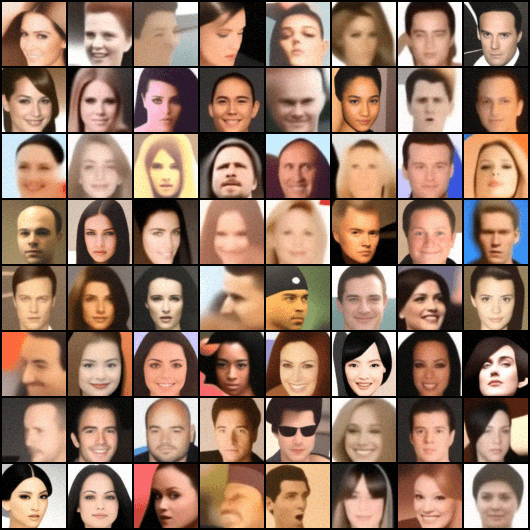}
  \end{minipage}
}
\subfigure[2-bit]{
    \begin{minipage}[b]{0.45\textwidth}
    \includegraphics[width=\textwidth]{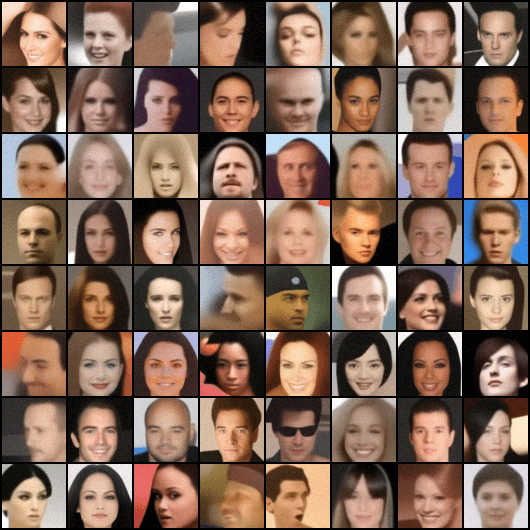}
    \end{minipage}
}
\subfigure[3-bit]{
  \begin{minipage}[b]{0.45\textwidth}
    \includegraphics[width=\textwidth]{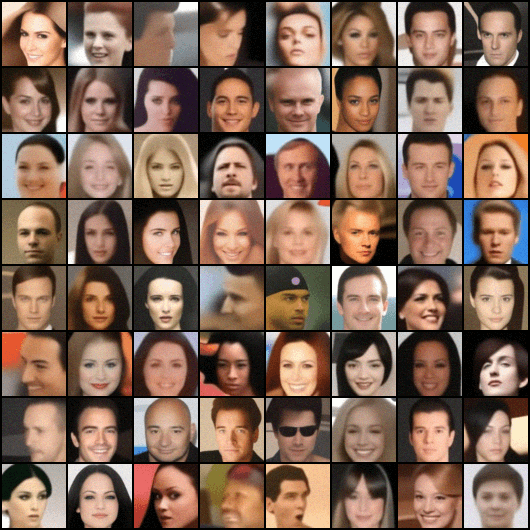}
  \end{minipage}
}
\caption{Results of QCS-SGM under 1-3 bit CS on CelebA for correlated $\mathbf{A}$ ($\rho=0.4$) when $M=4000, \sigma=0.1$.  }
\label{celeba-corr-qcs-sgm}
\end{figure*}

\begin{figure*}[htbp]
\centering
\subfigure[Truth]{
    \begin{minipage}[b]{0.45\textwidth}
    \includegraphics[width=\textwidth]{./figures/celeba_true_appendix.png}
    \end{minipage}
}
\subfigure[1-bit]{
  \begin{minipage}[b]{0.45\textwidth}
    \includegraphics[width=\textwidth]{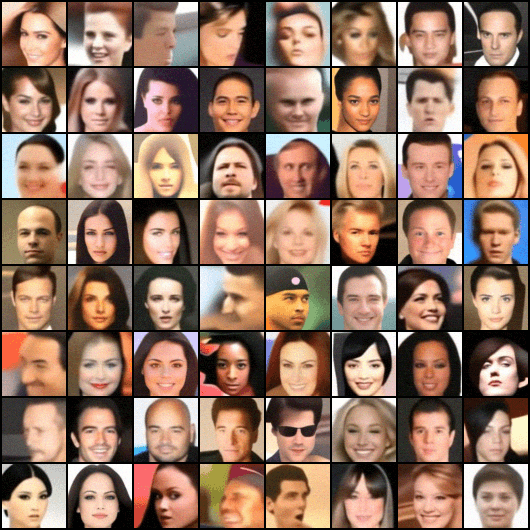}
  \end{minipage}
}
\subfigure[2-bit]{
    \begin{minipage}[b]{0.45\textwidth}
    \includegraphics[width=\textwidth]{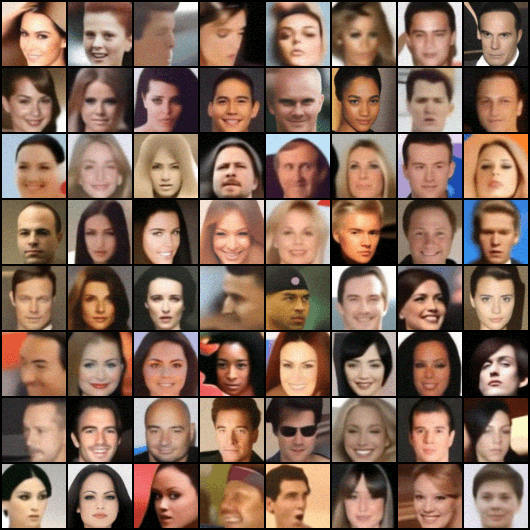}
    \end{minipage}
}
\subfigure[3-bit]{
  \begin{minipage}[b]{0.45\textwidth}
    \includegraphics[width=\textwidth]{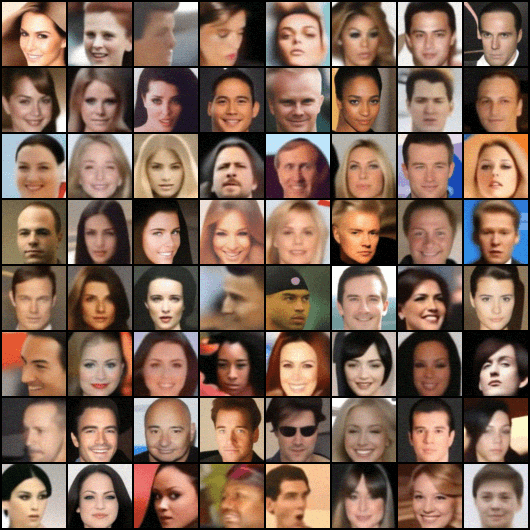}
  \end{minipage}
}
\caption{{Results of QCS-SGM+ (ours) under 1-3 bit CS on CelebA for correlated $\mathbf{A}$ ($\rho=0.4$) when $M=4000, \sigma=0.1$.   It can be seen that QCS-SGM+ outperforms QCS-SGM in Figure \ref{celeba-corr-qcs-sgm} remarkably.}}
\label{celeba-corr-qcs-sgm-plus}
\end{figure*}

In this section, we provide additional qualitative results in various settings, as shown in Figures \ref{mnist_compare_appendix} - \ref{celeba-corr-qcs-sgm-plus}. The results in Figures \ref{mnist_compare_appendix} - \ref{celeba-corr-qcs-sgm-plus} empirically demonstrate that our proposed QCS-SGM+ remarkably outperforms the original QCS-SGM for general matrices beyond mere row-orthogonality. 

\end{document}